\documentclass[10pt]{iopart}

\usepackage{iopams}
\usepackage[T1]{fontenc}
\usepackage[utf8]{inputenc}
\usepackage{graphicx}
\usepackage{tabularx}
\usepackage{color}
\usepackage{textcomp}
\usepackage{enumerate}
\usepackage{todonotes}
\usepackage{hyperref}
\usepackage{comment}
\usepackage{url}
\usepackage{caption}
\usepackage{iopams}
\usepackage{cite}
\usepackage{url}
\usepackage{color,soul}

\usepackage{xcolor}

\providecommand{\revision}[1]{{#1}}

\usepackage{todonotes}

\begin{document}

\title{A comparison of particle and fluid models for positive streamer discharges in air}
\author{Zhen Wang$^{1,2}$, Anbang Sun$^{1}$, Jannis Teunissen$^{2}$}
\address{$^1$State Key Laboratory of Electrical Insulation and Power Equipment, School of Electrical Engineering,
  Xi'an Jiaotong University, Xi'an, 710049, China,
  $^2$Centrum Wiskunde \& Informatica, Amsterdam, The Netherlands\\
	}
\ead{jannis.teunissen@cwi.nl, anbang.sun@xjtu.edu.cn}

\begin{abstract}
  Both fluid and particle models are commonly used to simulate streamer
  discharges. In this paper, we quantitatively study the agreement between these
  approaches for axisymmetric and 3D simulations of positive streamers in air.
  We use a drift-diffusion-reaction fluid model with the local field
  approximation and a PIC-MCC (particle-in-cell, Monte Carlo collision) particle
  model. The simulations are performed at 300 K and 1 bar in a 10 mm plate-plate
  gap with a 2 mm needle electrode. Applied voltages between 11.7 and 15.6
  kV are used, which correspond to background fields of about 15 to 20 kV/cm. Streamer properties like maximal electric field, head position and velocity are
  compared as a function of time or space.

  Our results show good agreement between the particle and fluid
  simulations, in contrast to some earlier comparisons that were carried out in
  1D or for negative streamers.
  % When compared at the same time, differences in
  % streamer length and maximal electric field are generally below 3\% and 6\%,
  % respectively.
  To quantify discrepancies between the models, we mainly look at streamer
  velocities as a function of streamer length. For the test cases considered
  here, the mean deviation in streamer velocity between the particle and fluid
  simulations is less than 4\%. % \jt{7.6\%}
  We study the effect of different types of transport data for the fluid model,
  and find that flux coefficients lead to good agreement whereas bulk
  coefficients do not. Furthermore, we find that with a two-term Boltzmann
  solver, data should be computed using a temporal growth model for the best agreement.
  The numerical convergence of the particle and fluid models is also studied. In
  fluid simulations the streamer velocity increases somewhat using finer grids,
  whereas the particle simulations are less sensitive to the grid.
  Photoionization is the dominant source of stochastic fluctuations in our
  simulations. When the same stochastic photoionization model is used, particle
  and fluid simulations exhibit similar fluctuations.
\end{abstract}

\maketitle

% Uncomment for two-column
\ioptwocol

\section{Introduction}

Streamer discharges are elongated conducting channels that typically appear when
an insulating medium is locally exposed to a field above its breakdown
value~\cite{nijdam2020physics}. Electric field enhancement around their tips
causes streamers to rapidly grow through electron impact ionization. Due to this
field enhancement, they can propagate into regions in which the background field
was initially below breakdown. Streamers occur in nature as
sprites~\cite{ebert2010review,Pasko_2013} and they are used in diverse applications such as
the production of radicals~\cite{kanazawa2011observation}, pollution
control~\cite{Kim_2004}, the treatment of liquids~\cite{Bruggeman_2016}, plasma
medicine~\cite{Keidar_2013}, and plasma
combustion~\cite{starikovskaia2014plasma}.

Over the last decades, numerical simulations have become a powerful tool to
study streamer physics and to explain experimental results. Two types of models
have commonly been used: particle models
(e.g.,~\cite{Rose_2011,teunissen20163d,Kolobov_2016,Levko_2017a,Stephens_2018a})
and fluid models
(e.g.,~\cite{Babaeva_2016,teunissen2017simulating,Plewa_2018,bagheri2019effect,Marskar_2019a,Starikovskiy_2020,Ono_2020}).

Particle models track the evolution of a large number of electrons, represented
as (super-)particles, and other relevant species. They can be used to study
stochastic phenomena such as electron runaway or discharge inception. Another
advantage of such models is that no assumptions on the EVDF (electron velocity
distribution function) need to be made. In fluid models all relevant species are
approximated by densities, which can greatly reduce computational costs. These
densities evolve due to fluxes and source terms, which are computed by making
certain assumptions about the EVDF. Common is the local field approximation, in
which it is assumed that electrons are instantaneously relaxed to the local
electric field. \revision{Higher-order fluid models can also be used~\cite{Dujko_2013,Garland_2017}. In principle, it would even be possible to solve the underlying spatio-temporal Boltzmann equation~\cite{Kortshagen_1996,Trunec_2006}, but this is at present computationally infeasible for multi-dimensional streamer simulations.}

\revision{Both particle models and fluid models with the local field
  approximation have frequently been used to study positive streamer discharges.
  Although it is well known that the local field approximation can lead to
  errors~\cite{Drallos_1995,Grubert_2009}, it is not clear how significant these
  errors are for the modeling of positive streamers.} The first goal of this
paper is therefore to study the agreement between particle and fluid simulations
of positive streamers in air, using both axisymmetric and 3D simulations. We use
a standard particle model of the PIC-MCC (particle-in-cell, Monte Carlo
collision) type and a standard drift-diffusion reactions fluid model with the
local field approximation.

When comparing models, it is important to have consistent input data computed
from the same electron-neutral cross sections. However, transport coefficients
for a fluid model can be computed with different types of Boltzmann solvers, and
both so-called \emph{flux} and \emph{bulk} coefficients can be computed. Bulk
coefficients describe the dynamics of a group of electrons, whereas flux
coefficients characterize the properties of individual electrons~\cite{Petrovic_2009a,Dujko_2013}. Although the
use of flux coefficients is \revision{generally
recommended for plasma modeling}~\cite{robson2005colloquium,Dujko_2013}, \revision{it is not fully clear how
the use of bulk data affects simulations of positive streamers~\cite{li2012comparison}}.
Furthermore, with a two-term Boltzmann solver it possible to use either a
spatial or temporal growth model, which lead to different transport
coefficients~\cite{hagelaar2005solving}. The second goal of this paper is
therefore to determine the most suitable type of transport data for use in fluid
simulations of positive streamers.

\textbf{Past work} Below, we briefly discuss some of the past work on the
comparison of particle and fluid models for streamer discharges.
In~\cite{li2012comparison}, four models were compared by simulating a short
negative streamer in 3D: a particle model, the `classical' fluid model with the
local field approximation, an extended fluid model, and a hybrid particle-fluid
model. These simulations were carried out in a $1.2$ mm gap using a background
electric field well above breakdown, without taking photoionization into
account. The classical fluid model here deviated from the other models in terms
of streamer velocity and shape, but this was probably due to an implementation
flaw that was later found. In~\cite{markosyan2015comparing} three plasma fluid models of different order
were compared against particle simulations for planar ionization waves, which
can be thought of as ``1D'' negative streamers. The classical fluid model was
found to give rather reasonable results, somewhat in contrast with the
conclusions of~\cite{Grubert_2009}. Finally, in \cite{bagheri2018comparison} six
streamer codes from different groups were benchmarked against each other, aiming
towards model verification. All codes implemented the classical fluid model, and
three test cases with positive streamers were considered. Good agreement was
found on sufficiently fine grids, and with corresponding small time
steps.

Particle and fluid models have also been compared for other types of discharges. In~\cite{kim2005particle} particle, fluid and hybrid
models were benchmarked against each other and against experimental data. This
review paper focused on applications related to plasma display panels,
capacitively coupled plasmas and inductivetly coupled plasmas. The authors
conclude that ``\emph{Excellent agreement can be found in these systems when the
  correct model is used for the simulation. Choosing the right model requires an
  understanding of the capabilities and limitations of the models and of the
  main physics governing a particular discharge.}''. Furthermore, in~\cite{Lee_2006}, particle and fluid models were compared for capacitively and inductively coupled argon-oxygen plasmas, in~\cite{Hong_2008} they were compared for atmospheric pressure
helium microdischarges, and in~\cite{Becker_2017a} they were extensively compared for low-pressure ccrf discharges.

% A few further studies in comparing different photoionization models were also carried out, see \cite{bourdon2007efficient,bagheri2019effect}.
% In total, no detailed comparison among multi-dimension models with the same framework is currently made.

% The results showed that the hybrid model overcame super-particle artifacts and
% preserved real physical fluctuations, while the extended fluid model gave a good
% approximation until front destabilization, and the classical fluid model behaves
% quantitatively low in propagation speeds, ionization densities and field
% enhancement.
% Chao Li $et$ $al$ validated their 3D spatial coupling hybrid model through comparing it with both particle and fluid models, and the comparison objects were negative streamers in air without photoionization below the breakdown voltage.

In contrast to the above work, we here compare multidimensional particle and fluid simulations of positive streamers in air, propagating in background fields below breakdown, including photoionization.
The paper is organized as follows. In section \ref{sec:model-description}, the
particle and fluid models are described as well as the simulation conditions. In
section \ref{subsec:15-comparison}, we first compare axisymmetric and 3D
particle and fluid simulations of positive streamers in air, after which the
influence of transport data is studied with axisymmetric fluid simulations in
section \ref{sec:TD-data}. We then investigate the numerical convergence of both
types of models in section \ref{sec:mesh-refin-numer}, and we determine the
dominant source of stochastic fluctuations in the particle simulations in
section \ref{sec:role-stoch-effects}. Finally, in section
\ref{sec:results-at-different}, the models are again compared for different
applied voltages.

\section{Model description}
\label{sec:model-description}

The particle and fluid models are briefly introduced below, after which the
simulation conditions, photoionization and the adaptive mesh are described. We use both
axisymmetric and 3D models. For brevity, axisymmetric models will sometimes be referred to as ``2D''.

Due to the short time scales considered in this paper, ions are assumed to be immobile in both the particle and fluid models.
Information about the computational cost of simulations is given in
\ref{sec:computational-costs}.

% , where PIC  means charged particles such as electrons and ions are mapped into numerical grids

% Since every single electron can be traced by recording its motion and location, one can use this model to study microscopic properties of streamers.

% Due to the large number of electrons in streamer,which is normally 10$^7$—10$^9$ during inception stage and continues increasing \cite{meek1940theory, montijn2006diffusion}, particle models are of high computational costs.

\subsection{Particle (PIC-MCC) model}
\label{sec:particle-model}

We use a PIC-MCC (particle-in-cell, Monte-Carlo Collision) model that combines the particle model described in~\cite{teunissen20163d} with the Afivo AMR (adaptive mesh refinement) framework described in \cite{teunissen2018afivo}. Electrons are tracked as particles, ions as densities, and neutrals as a background that electrons stochastically collide with. Below, we briefly introduce the model's main components.

\subsubsection{Particle mover and collisions}
\label{sec:superparticles}

The coordinates $\vec{x}$ and velocities $\vec{v}$ of simulated electrons are advanced with the `velocity Verlet' scheme described in~\cite{teunissen20163d}:
\begin{eqnarray}
  \vec{x}(t + \Delta t) &= \vec{x}(t) + \Delta t \, \vec{v}(t) + \frac{1}{2} \Delta t^2 \, \vec{a}(t),\label{eq:mover-x}\\
  \vec{v}(t + \Delta t) &= \vec{v}(t) + \frac{1}{2} \left[\vec{a}(t) + \vec{a}(t+\Delta t) \right],\label{eq:mover-v}
\end{eqnarray}
where $\vec{a} = -(e/m_e) \vec{E}$ is the acceleration due to the electric field $\vec{E}$, and $e$ is the elementary charge and $m_e$ the electron mass.

In axisymmetric simulations particles are evolved as in a 3D Cartesian geometry. However, their acceleration, which is due to an axisymmetric field, is projected onto a radial and axial components before it is used. The acceleration in equation (\ref{eq:mover-x}) is then given by $\vec{a} = (a_r \, x/r, a_r \, y/r, a_z)$, where $x$ and $y$ denote the two (3D) particle coordinates corresponding to the radial direction and $r = \sqrt{x^2 + y^2}$. In equation (\ref{eq:mover-v}), the radial velocity is updated as
\begin{equation*}
  \vec{v}_{x,y}(t + \Delta t) = \vec{v}_{x,y}(t) +  \frac{1}{2} \hat{r} \left[a_r(t) + a_r(t+\Delta t) \right],
\end{equation*}
where $\hat{r} = (x, y)/r$.

Electron-neutral collisions are handled with the null-collision method~\cite{Koura_1986,teunissen20163d},  using collision rates calculated from cross section input data.

\subsubsection{Super-particles}
\label{sec:superparticles}

Due to the large number of electrons in a streamer discharge, it is generally not feasible to simulate all electrons individually. Instead, so-called ``super-particles'' are used, whose weights $w_i$ determine how many physical particles they represent~\cite{Hockney_1988}. The procedure followed here is similar to that in~\cite{teunissen20163d}. A parameter $N_{ppc}$ controls the `desired' number of particles per grid cell. We use $N_{ppc} = 75$, except for section \ref{sec:role-stoch-effects}, in which it is varied. The desired particle weights $\omega$ are then determined as
\begin{eqnarray}
  \omega =  n_e\times \Delta V / N_{ppc}
  \label{eq:weight-calculation-superparticle}
\end{eqnarray}
where $n_e$ is the electron density in a cell and $\Delta V$ the cell volume. Furthermore, the minimum particle weight is $\omega_\mathrm{min} = 1$.

Particle weights are updated when the number of simulations particles has grown by a factor of $1.25$, after the AMR mesh has changed, or after 10 time steps in axisymmetric simulations (see below). The particles in a cell for which $w_i < (2/3) \times \omega$ are merged, by combining two such particles that are close in energy into one with the sum of the original weights. The coordinates and velocity of the merged particle are randomly selected from one of the original particles, see~\cite{Teunissen_2014a}. Particles are split when $w_i > (3/2) \times \omega$. Their weight is then halved after which identical copies of these particles are added, which will soon deviate from them due to the random collisions.

\begin{figure}
  \centering
  \includegraphics[width=\linewidth]{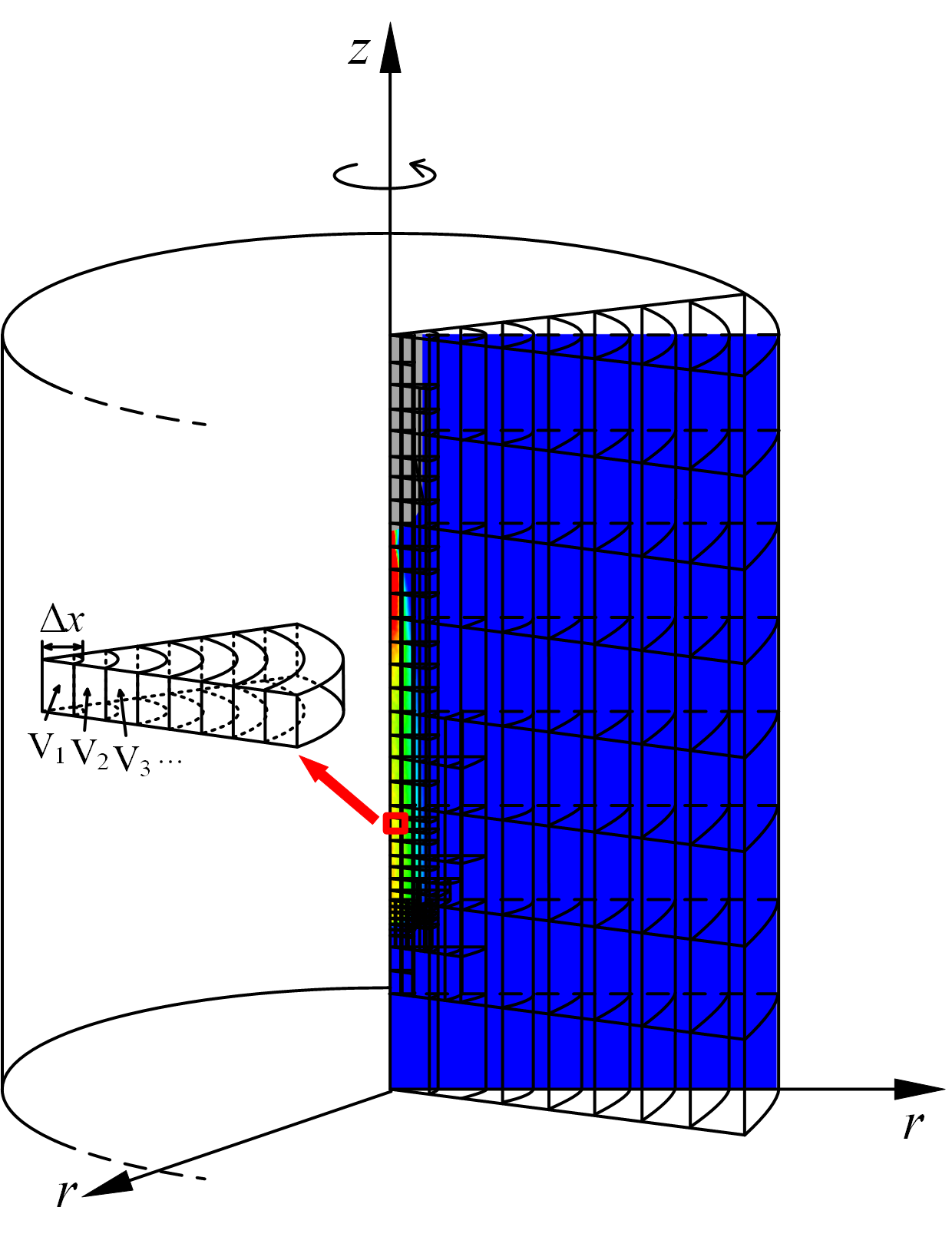}
  \caption{Illustration of the adaptive mesh in an axisymmetric streamer
    simulation, also showing the electron density. The grid is coarser than in
    an actual simulation. The enlarged view illustrates the relation between
    cell volumes and radius.}
  \label{fig:cylindrical-coordinates}
\end{figure}

In axisymmetric simulations particle weights are updated every ten
time steps to reduce fluctuations near the axis. Figure \ref{fig:cylindrical-coordinates}
illustrates an axisymmetric mesh in which cell
volumes depend on the radial coordinate as
$\Delta V = 2\pi r \Delta r \Delta z$. % We use a cell-centered discretization, in
% which the cell $i$ is centered at $(i - 0.5) \Delta r$ for $i \in 1, 2, \dots$,
% so that the ratio of cell volumes is $\Delta V_i/\Delta V_1 = 2i - 1$ if they
% are at the same refinement level.
Cells with small volumes contain fewer
physical electrons, and because the minimal super-particle weight is one,
stochastic fluctuations in such cells are larger. Furthermore, super-particle
weights given by equation (\ref{eq:weight-calculation-superparticle}) are
proportional to the cell volume. Particles with high weights can therefore cause
significant fluctuations when they move towards the axis. We update the particle
weights more frequently in axisymmetric simulations to limit these fluctuations.

\subsubsection{Mapping particles to the grid}
\label{sec:mapping-to-grid}

Particles are mapped to grid densities using a standard bilinear or trilinear weighting scheme, as in \cite{teunissen20163d}. Near refinement boundaries, the mapping is locally changed to the ``nearest grid point'' (NGP) scheme, to ensure that particle densities are conserved. In axisymmetric coordinates the mapping is also done using bilinear weighting, but the radial variation in cell volumes is taken into account. The interpolation of the electric field from the grid to particles is done using standard bilinear (2D) or trilinear (3D) interpolation. Note that there are more advanced weighting schemes for handling axisymmetric coordinates systems~\cite{Araki_2014}, but these approaches are challenging to combine with the cell-centered AMR used in our models.

\subsubsection{Temporal discretization}
\label{sec:time-step-pic}

We use the following CFL-like condition
\begin{eqnarray}
  \Delta {t_\mathrm{cfl}} \, \tilde{v}_{\mathrm{max}} \leq 0.5 \times \Delta {x_{\min }},
  % \label{eq:dt-cfl}
\end{eqnarray}
where $\Delta {x_{\min }}$ indicates the minimum grid spacing, and $\tilde{v}_{\mathrm{max}}$ is an estimate of the particle velocity at the 90\%-quantile. This prevents a few fast particles from affecting $\Delta {t_\mathrm{cfl}}$.

Another time step constraint is the Maxwell time, also known as the dielectric relaxation time, which is a typical time scale for electric screening:
\begin{equation}
  \Delta {t_\mathrm{drt}} = \varepsilon _0/\left(e n_{e,\mathrm{max}} \, \mu _e\right),
  \label{eq:dt-drt}
\end{equation}
where $\varepsilon _0$ is the dielectric permittivity, $n_{e,\mathrm{max}}$ is the maximal electron density, and $\mu _e$ the electron mobility, as determined in the local field approximation (see section \ref{subsec:fluid-input-data}).

The actual time step is then the minimum of $\Delta {t_\mathrm{cfl}}$ and $\Delta {t_\mathrm{drt}}$, and it is furthermore adjusted such that the number of electrons does not grow by more than 20\% between time steps.

% Meanwhile, the increasing rate of electron numbers should be limited in a certain range:
% \begin{eqnarray}
% \Delta {t_e} = \left\{ \begin{array}{l}
% \min (2\Delta {t_{n - 1}},\Delta {t_{n - 1}} \times {({\sigma _{e,\max }}/({\sigma _e}{\rm{ + R}}))^{0.1}},{\sigma _e} < 0.5{\sigma _{e,\max }}\\
% \Delta {t_{n - 1}},0.5{\sigma _{e,\max }} \le {\sigma _e} \le {\sigma _{e,\max }}\\
% \Delta {t_{_{n - 1}}} \times {\sigma _{e,\max }}/{\sigma _e},{\sigma _e} > {\sigma _{e,\max }}
% \end{array} \right.
%   \label{eq:dt-drt}
% \end{eqnarray}
% where $t_{n-1}$ is the time step of last iteration, $\sigma_{e}$ and $\sigma_{e,max}$ is the increasing rate of electron numbers and its upper limit, here we set as 20\%.

\subsubsection{Input data}
\label{sec:input-data}

We use Phelps' cross sections for N$_2$ and O$_2$ \cite{phelps1985anisotropic,Phelps-database}. These cross sections contain a so-called \emph{effective} momentum transfer cross sections, which account for the combined effect of elastic and inelastic processes~\cite{Pitchford_1982}. To use them in particle simulations, we convert them to elastic cross sections by subtracting the sum of the inelastic cross sections. This is an approximate procedure, but the resulting cross sections are suitable for a model comparison.

\subsection{Fluid model}
\label{sec:fluid-model}

We use a drift-diffusion-reaction fluid model with the local field approximation, as implemented in \cite{teunissen2017simulating}. The electron density $n_e$ evolves in time as
 \begin{equation}
   \partial_t n_e =  \nabla \cdot (n_e\mu _e\mathbf{E} + D_e\nabla n_e) + S_R + S_{ph}
   \label{eq:fluid-model}
 \end{equation}
 where $\mu_e$ and $D_e$ indicate the electron mobility and the diffusion
 coefficient, $S_{ph}$ is the non-local photoionization source term (see section
 \ref{sec:photoionization}), and $S_R$ is a source term due to the following
 ionization and attachment reactions:
 \begin{eqnarray*}
   \label{eq:reaction-list}
   \textrm{e} + \textrm{N}_2 &\longrightarrow 2\textrm{e} + \textrm{N}_2^+\textrm{ (15.60 eV)}\\
    \textrm{e} + \textrm{N}_2 &\longrightarrow 2\textrm{e} + \textrm{N}_2^+\textrm{ (18.80 eV)}\\
    \textrm{e} + \textrm{O}_2 &\longrightarrow 2\textrm{e} + \textrm{O}_2^+\\
    \textrm{e} + 2\textrm{O}_2 &\longrightarrow \textrm{O}_2^- + \textrm{O}_2\\
    \textrm{e} + \textrm{O}_2 &\longrightarrow \textrm{O}^- + \textrm{O}
 \end{eqnarray*}
Ion densities also change due to the above reactions.
Transport coefficients and reaction rates are determined using the local field approximation, see section \ref{subsec:fluid-input-data}.

\subsubsection{Time integration}
\label{sec:time-step-fluid}

Advective electron fluxes are computed using the Koren flux limiter~\cite{Koren_1993} and diffusive fluxes using central differences, see \cite{teunissen2017simulating} for details.
Time integration is performed with Heun's method, a second order accurate explicit Runge-Kutta scheme. Time steps are limited according to the following restrictions:
\begin{eqnarray*}
  \label{eq:dt-fluid}
  &\Delta t_\mathrm{cfl} \left(\frac{2 N_\mathrm{dim}}{\Delta x^2} + \sum \frac{v_i}{\Delta x} \right) \leq 0.5,\\
  &\Delta t_\mathrm{drt} \left(e n_e \mu_e  / \varepsilon_0\right) \leq 1,\\
  &\Delta t = 0.9 \times \min(\Delta t_\mathrm{drt}, \Delta t_\mathrm{cfl}),
\end{eqnarray*}
where $\Delta t_\mathrm{cfl}$ corresponds to a CFL condition (including diffusion), $\Delta t_\mathrm{drt}$ corresponds to the dielectric relaxation time (as in equation (\ref{eq:dt-drt})), $N_\mathrm{dim}$ is the dimensionality of the simulation, $\Delta t$ is the actual time step used, and $\Delta x$ stands for the grid spacing, which is here equal in all directions.

\subsubsection{Input data}
\label{subsec:fluid-input-data}

The fluid model requires tables of transport and reaction data versus electric
field strength as input. We use two methods to compute such data from the cross
sections that were also used for the PIC model, see section
\ref{sec:input-data}. The first is BOLSIG+, which is a widely used two-term
Boltzmann solver~\cite{hagelaar2005solving,bolsig2019}. When the electron
velocity distribution is strongly anisotropic, for example in high electric
fields, the use of the two term approximation can introduce
errors~\cite{bankovic2012approximations}. The second method we use is a Monte
Carlo swarm code \url{gitlab.com/MD-CWI-NL/particle_swarm}, similar to
e.g.~\cite{biagi1999monte,rabie2016methes}. The basic idea of this approach is
to trace electrons in a uniform field, from which transport and reactions
coefficients can be obtained.

With the Monte Carlo method we compute both so-called \emph{flux} and
\emph{bulk} coefficients~\cite{Petrovic_2009a,Dujko_2013}. Flux data describes
the average properties of individual electrons, whereas bulk data describes
average properties of a group of electrons, taking \revision{non-conservative
  collisions such as ionization and attachment into account. Consider for
  example a group of electrons, which changes in size due to non-conservative
  collisions. The bulk drift velocity then describes the average velocity of the
  center of mass of this group, whereas the flux drift velocity describes the
  average velocity of individual electrons. These two definitions differ when
  the probability of non-conservative collisions is not uniform in space, which
  causes motion of the center of mass.}

One of the main
differences \revision{between bulk and flux data} is that in high fields the bulk mobility is higher than the flux
mobility, as shown in figure \ref{fig:transport_value_linear}. Unless mentioned
otherwise, the fluid simulations presented in this paper use Monte Carlo flux
data.

% reflected in their mobilities $\mu$ and diffusion coefficients $D$, since the bulk mobility represents the average location changing rate for the mass centers of each electron under an external electric fields, while the flux mobility represents the changing rate of the mass center location for total electron group.

\subsection{Computational domain and initial conditions}
\label{subsec:initial-condition}

Simulations are performed in artificial air, containing 80$\%$ N$_2$ and 20$\%$
O$_2$, at $p$ = 1 bar and $T$ = 300 K. \revision{We will give electric fields in
  units of kV/cm. With a gas number density of $N = 2.414\times 10^{25}$
  m$^{-3}$, assuming the ideal gas law, 1 kV/cm corresponds to about $4.14$ Td
  (Townsends).}

\begin{figure}
  \centering
  \includegraphics[width=\linewidth]{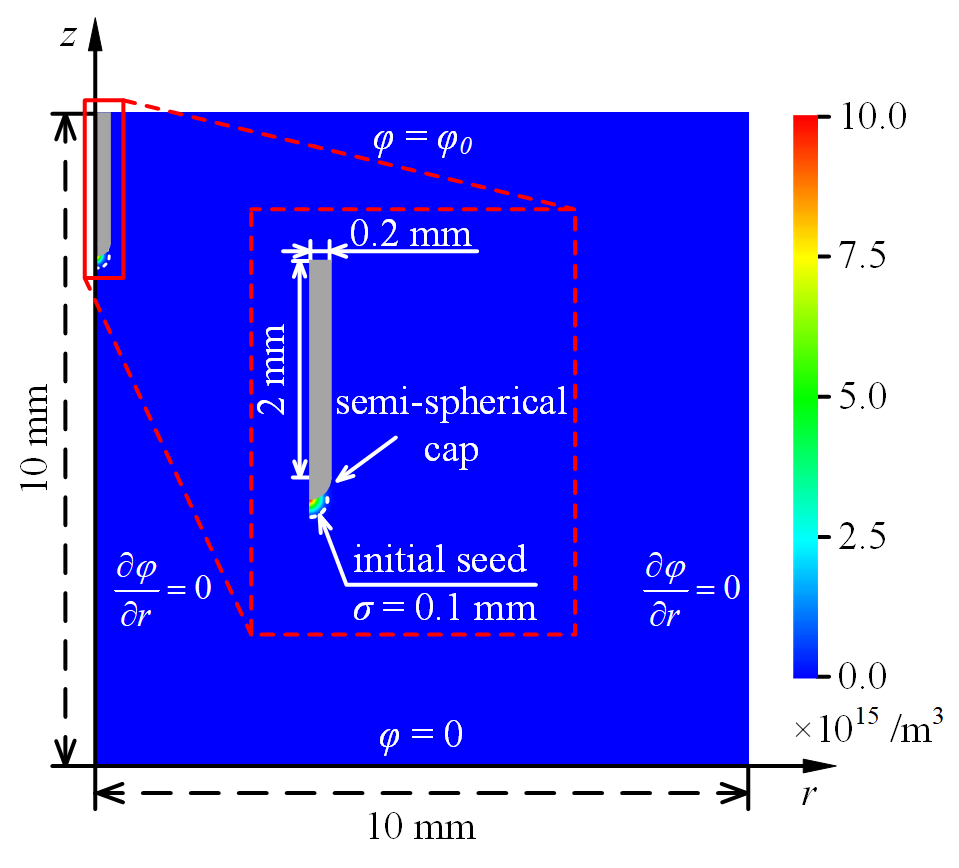}
  \caption{Schematic view of the axisymmetric computational domain used for
    both particle and fluid models, showing the initial electron density and the
    electrode. 3D simulations are performed in a similarly sized domain,
    measuring 20 mm $\times$ 20 mm $\times$ 10 mm.}
  \label{fig:initial-condition}
\end{figure}

The computational domain used for the comparison of cylindrical models is shown in figure \ref{fig:initial-condition}. It measures 10 mm in both the axial and radial directions. For the 3D Cartesian simulations, a similar domain of 20 mm $\times$ 20 mm $\times$ 10 mm is used. A rod-shaped electrode with a semi-spherical cap is placed at center of the domain. This electrode is 2.0 mm long and has a radius of 0.2 mm.

For the electric potential, Dirichlet boundary conditions are applied to the lower and upper domain boundaries, and a homogeneous Neumann boundary condition is applied on the other boundaries. In terms of electrostatics the axisymmetric and 3D simulations are not fully equivalent, because of the different geometry in which these boundary conditions are applied.

For the electron density homogeneous Neumann conditions are applied on all domain boundaries, except for the rod electrode. The electrode absorbs electrons but does not emit them. Since a positive voltage is applied on this electrode, secondary electron emission was not taken into account.

There is initially no background ionization besides an electrically neutral plasma seed that is placed at the tip of the electrode. This seeds helps to start discharges in almost the same way in particle and fluid models. The electron and positive ion densities are given by a Gaussian distribution:
\begin{equation}
  n_i(\mathbf{r})=n_e(\mathbf{r})= 10^{16}\,\mathrm{m}^{-3}
  \exp\left[\frac{(\mathbf{r}-\mathbf{r_0})^2}{(0.1 \, \mathrm{mm})^2}\right],
  \label{eq:gaussian-distribution}
\end{equation}
where $\mathbf{r_0}$ is the location of the tip of the electrode, which is at $z \approx 7.8 \, \textrm{mm}$.

In particle simulations, these initial densities are converted to $N = \lfloor n_e \Delta V\rfloor$ simulation particles per cell, each with a weight of one.
A uniform $[0, 1)$ random
number is compared to the remainder to determine whether to add one more
particle. The goal of this sampling is to reduce stochastic fluctuations in
the initial conditions. The initial particles are spread uniformly within each grid cell, and they initially have zero kinetic energy.

% If $N > 0$, particle weights are determined by rounding $n_e \Delta V / N$ to an integer.

% \begin{eqnarray}
%   &N_{cell}(i) = \min [\frac{n_e(i)\times V_i}{\omega_{min}}, N_{ppc}]\\
%   &\omega(i) =  \frac{n_e(i)\times V_i}{N_{cell}(i)}
%     \label{eq:weight-calculation}
% \end{eqnarray}
% where $N_{cell}(i)$ is the particle number in cell $i$, $n_e(i)$ is the electron density of cell $i$, it is calculated from the interpolation of a desired density distribution, a detailed discuss can be found in section 2.3. $\omega_{min}$ is the required minimum weight for particles, and $N_{ppc}$ indicates the minimal particle number per cell, this restriction is imposed in whole simulation process to avoid the sudden density drop caused by the small cell volumes near axis, here it is set as 75 for all particle models in model comparison sections (except for section 3.3).

\subsection{Photoionization}
\label{sec:photoionization}

For streamers in air, photoionization is typically the main source of free electrons. The process results from the interaction between an oxygen molecule and an UV photon emitted from an excited nitrogen molecule. We use Zheleznyak's model for photoionization in air~\cite{zhelezniak1982photoionization} and compute photoionization as in \cite{bagheri2019effect}. Assuming that ionizing photons do not scatter and their direction is isotropically distributed, the photoionization source term in equation (\ref{eq:fluid-model}) is given by:
\begin{equation}
  {S_{ph}}(\mathbf{r}) =
  \int{\frac{I(r')f(\left|{\mathbf{r}-\mathbf{r'}}\right|)}{4\pi\left|{\mathbf{r}-\mathbf{r'}}\right|^2} d^3r'},
     \label{eq:photoionization}
\end{equation}
where $f(r)$ is the photon absorption function and $I(\mathbf{r})$ is the source of ionizing photons, which is proportional to the electron impact ionization source term $S_i$:
\begin{equation}
  I(\mathbf{r}) = \frac{p_q}{p+p_q}\xi S_i,
  \label{eq:UV-source-term}
\end{equation}
where $p$ is the gas pressure, $p_q = 40$ mbar is a quenching pressure. For simplicity, we use a constant proportionality factor $\xi = 0.075$, except for section \ref{sec:role-stoch-effects}, in which $\xi$ is varied.

We solve equation (\ref{eq:photoionization}) in two ways in this paper. For fluid simulations, we use the so-called Helmholtz expansion~\cite{bourdon2007efficient,Luque_2007}.
By approximating the absorption function, the integral in equation (\ref{eq:photoionization}) can be turned into multiple Helmholtz equations that can be solved by fast elliptic solvers. We use Bourdon's three-term expansion, as described in~\cite{bourdon2007efficient} and appendix A of \cite{bagheri2018comparison}.

For particle simulations, we use a discrete Monte Carlo photoionization model as described in~\cite{Chanrion_2008,bagheri2019effect}. With this model stochastic effects due to discrete single photons are simulated. The basic idea is to stochastically sample the generated photons, their directions, and their travel distances. We also use this approach for the fluid simulations in section \ref{sec:role-stoch-effects}, see~\cite{bagheri2019effect} and chapter 11 of \cite{teunissen20153dsimulations} for details.

These two approaches for photoionization differ not only in terms of stochastic effects. Because of the way the absorption function is approximated in the Helmholtz approach, the number of ionizing photons produced and their absorption profile will also be somewhat different. However, such small differences in photoionization usually have only a minor effect on discharge development, as confirmed by our results in section \ref{sec:results}.

\subsection{Afivo AMR framework}
\label{sec:afivo-amr-framework}

The open-source Afivo framework \cite{teunissen2018afivo} is used in both particle and fluid models to provide adaptive mesh refinment (AMR) and a parallel multigrid solver.
Adaptive mesh refinement (AMR) is used for computational efficiency, based on the following criteria~\cite{teunissen2017simulating}:
\begin{itemize}
\item refine if $\alpha(E) \Delta x > {c_0}$,
\item de-refine if $\alpha (E) \Delta x < 0.125 {c_0}$, but only if $\Delta x$ is smaller than 10 $\mu$m.
\end{itemize}
Here $\Delta x$ is the grid spacing, which is equal in all directions,
$\alpha(E)$ is the field-dependent ionization coefficient, and $c_0$ is a constant. Furthermore, the grid spacing is bound by
$\Delta x \leq 0.4$ mm. For 3D particle simulations we use $c_0 = 1.0$ and for
all other simulations $c_0 = 0.8$. Slightly less refinement is used for the 3D
particle simulations because of their large computational cost, see
\ref{sec:computational-costs}. With these values for $c_0$ numerical convergence
errors are reasonably small, as discussed in section \ref{sec:mesh-refin-numer}.

% To obtain results with relatively high resolution, we use the same refinement equation but with a more strict parameter, which is
% $\Delta$x$<$0.8$\alpha(E)$, where $\alpha$ is the field-dependent ionization coefficient, $E$ is the electric field strength in V$\cdot$m$^{-1}$, and $\Delta$x is the maximal value of grid side length in all directions.

The geometric multigrid solver in Afivo~\cite{teunissen2018afivo} is used to
efficiently solve Poisson's equation $\nabla^2 \phi = \rho/\varepsilon_0$, where
$\phi$ is the electric potential, $\varepsilon_0$ the permittivity of vacuum and
$\rho$ the space charge density. Electrostatic fields are then computed as
$\vec{E} = - \nabla \phi$. The same type of multigrid solver is also used to
solve the Helmholtz equations for photoionization. To include a needle
electrode, we set the applied potential as a boundary condition at the electrode
surface. This was implemented by modifying the multigrid methods using a
level-set function.

% TODO: After the mapping procedure, the electric potential can be calculated from solving Poisson's equation.

\section{Results}
\label{sec:results}

\subsection{Axisymmetric and 3D results}
\label{subsec:15-comparison}

\begin{figure*}
    \centering
    \includegraphics[width=\linewidth]{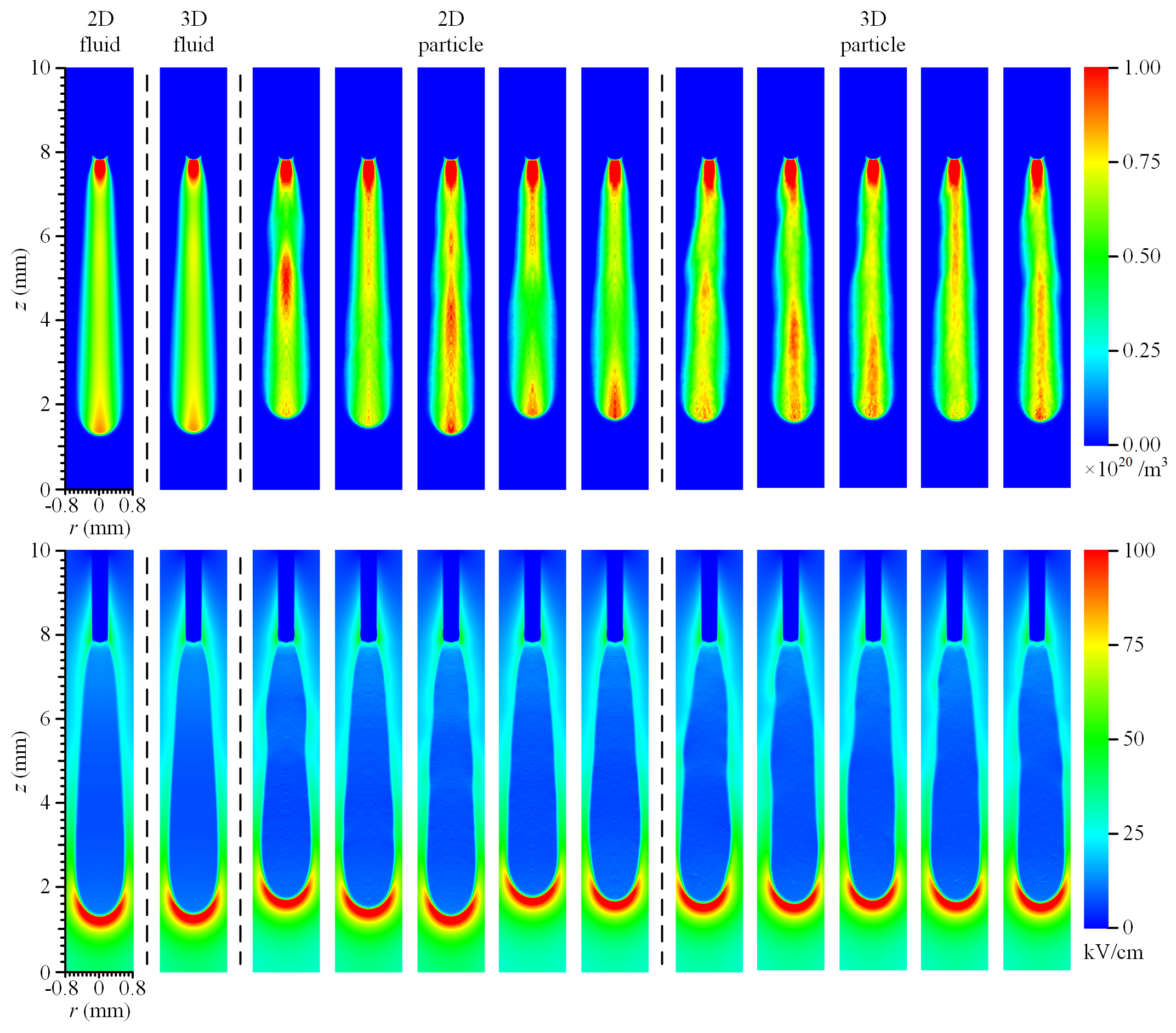}
    \caption{The electron densities and electric fields at $t$ = 10 ns for fluid
      and particle models at an applied voltage of 11.70 kV. The axisymmetric
      results are mirrored in the symmetry axis. For the 3D simulations cross
      sections are shown. Multiple runs are shown for the stochastic particle
      simulations. For the fluid simulations Monte Carlo flux transport data was
      used.}
    \label{fig:15-results}
  \end{figure*}

In this section, we compare axisymmetric and 3D particle and fluid models, using
the computational configuration and initial condition described in section
\ref{subsec:initial-condition}. A voltage of $\phi$ = 11.70 kV is applied, which
results in a background field of around 15 kV/cm; about half the breakdown field
of air.

Photoionization in the fluid model is here computed with the Helmholtz
approximation, whereas the particle model uses a Monte Carlo scheme with
discrete photons. To account for stochastic fluctuations, ten runs of the 2D
cylindrical and 3D particle models are performed, of which five are shown. For the fluid simulations flux transport data from Monte Carlo swarms is used, see section
\ref{subsec:fluid-input-data}.

Figure \ref{fig:15-results} shows the electron densities and electric fields for
the different models at $t = 10 \, \textrm{ns}$. The electric field and electron density profiles are similar for all cases, and the streamer head positions are in good agreement, with deviations in streamer length below 5\%. % \jt{4.7\%}
Streamer head positions in all models at $t = 3, 6, 9$~ns are given in table \ref{tab:head-position}.

\begin{table}
  \centering
  \begin{tabular}{lllll}
    Model      & Data      & $z$ (3 ns) & $z$ (6 ns) & $z$ (9 ns)   \\
    \hline
    PIC-2D     & -         & 6.45       & 4.74       & 2.41       \\
    PIC-3D     & -         & 6.45       & 4.71       & 2.39       \\
    fluid-2D   & flux      & 6.40       & 4.64       & 2.31       \\
    fluid-3D   & flux      & 6.42       & 4.66       & 2.35       \\
    \hline
    fluid-2D   & B+ temp.  & 6.42       & 4.71       & 2.47       \\
    fluid-2D   & B+ spat.  & 6.72       & 5.46       & 3.87       \\
    fluid-2D   & bulk-a    & 6.19       & 4.13       & 1.34       \\
    fluid-2D   & bulk-b    & 6.54       & 5.04       & 3.12
    % fluid-3D & BOLSIG+ ( & \jt{TODO?} & \jt{TODO?} & \jt{TODO?} \\
    % fluid-3D & BOLSIG+   & \jt{TODO?} & \jt{TODO?} & \jt{TODO?} \\
  \end{tabular}
  \caption{Streamer head position ($z$, in mm) at 3, 6 and 9 ns in different simulations, using an applied voltage of 11.7 kV. The bottom part of the
    table gives results for different types of transport data, see section
    \ref{sec:TD-data}. Here ``B+ (temp.)'' and ``B+ (spat.)'' respectively refer
    to flux data computed with BOLSIG+ using temporal growth and spatial growth
    models, and ``bulk(a)'' and ``bulk(b)'' refer to two types of bulk
    coefficients.}
  \label{tab:head-position}
\end{table}

With the axisymmetric particle model stochastic fluctuations are visible in the
streamer radius and the electron densities. As discussed in section
\ref{sec:role-stoch-effects} this is mainly due to the stochastic
photoionization used in the particle simulations. Streamers appear to propagate somewhat slower due to these fluctuations. In 3D similar
fluctuations are present, but the streamers can now move slightly off axis. The 3D particle model can in principle capture realistic stochastic fluctuations, but
only if single electrons are used instead of super-particles. This is
computationally not feasible for the simulations performed here.

Stochastic fluctuations are not present in the fluid simulations, in which the electron densities and electric fields evolve smoothly in time. % Streamer properties are affected by the used transport data, which is explored in more detail in section \ref{sec:TD-data}.
The results of the cylindrical and 3D fluid models are nearly identical. Small differences can occur because the computational domains correspond to a rectangle and a cylinder, which means that the applied boundary conditions are not equivalent. Furthermore, the numerical grids and operators are also slightly
different in these two geometries.

\begin{figure}
    \centering
    \includegraphics[width=\linewidth]{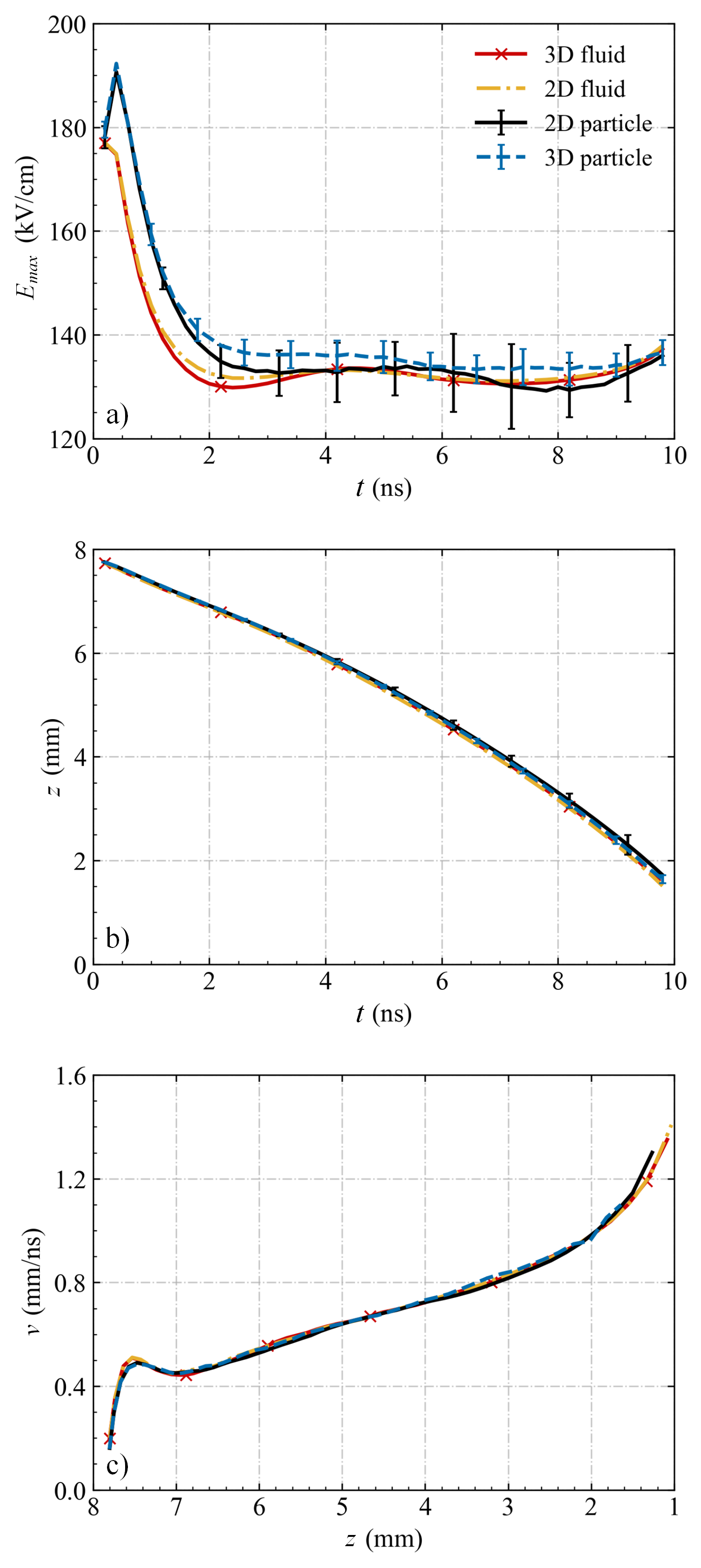}
    \caption{Comparison between axisymmetric and 3D particle and fluid
      simulations at an applied voltage of 11.70 kV. From top to bottom: maximal
      electric field versus time, streamer head position versus time and front velocity versus streamer position. For the stochastic particle simulations the
      average of ten runs is shown, and the error bars indicate $\pm$ one standard deviation.}
    \label{fig:15-comparison}
\end{figure}

To more quantitatively analyze the differences between models, the maximal field $E_{max}$, the streamer head position $z$ and streamer velocity $v$ are shown in figure \ref{fig:15-comparison}. The streamer head position is defined as the $z$-coordinate where the electric field is maximal. The velocity is shown versus streamer position, otherwise initial differences grow larger over time even if models agree well later on. The streamer velocity is computed as the numerical derivative of the streamer head position, which amplifies fluctuations. We use a second order Savitzky--Golay filter of width five to compute a smoothed velocity from the position versus time data.
For the stochastic particle simulations the average of ten runs is shown.

The maximal electric field follows
a similar trend in all models, with first a field of about 180 kV/cm and then a
relaxation towards a field of about 130--135 kV/cm as the streamers propagate
across the gap. When the streamers approach the grounded electrode the maximal
field increases again, because the available voltage difference is compressed in
a small region.

The peak electric fields during inception differ somewhat, with the particle
model having the highest peak at about 190 kV/cm whereas it is about 180 kV/cm
for both fluid models. The relaxation of this peak electric field occurs about
0.4 ns earlier in the fluid model. The main reason for this is that near the
electrode, the degree of ionization in the streamer channel is somewhat higher
in the particle simulations, which initially leads to stronger field
enhancement.

\begin{figure}
    \centering
    \includegraphics[width=\linewidth]{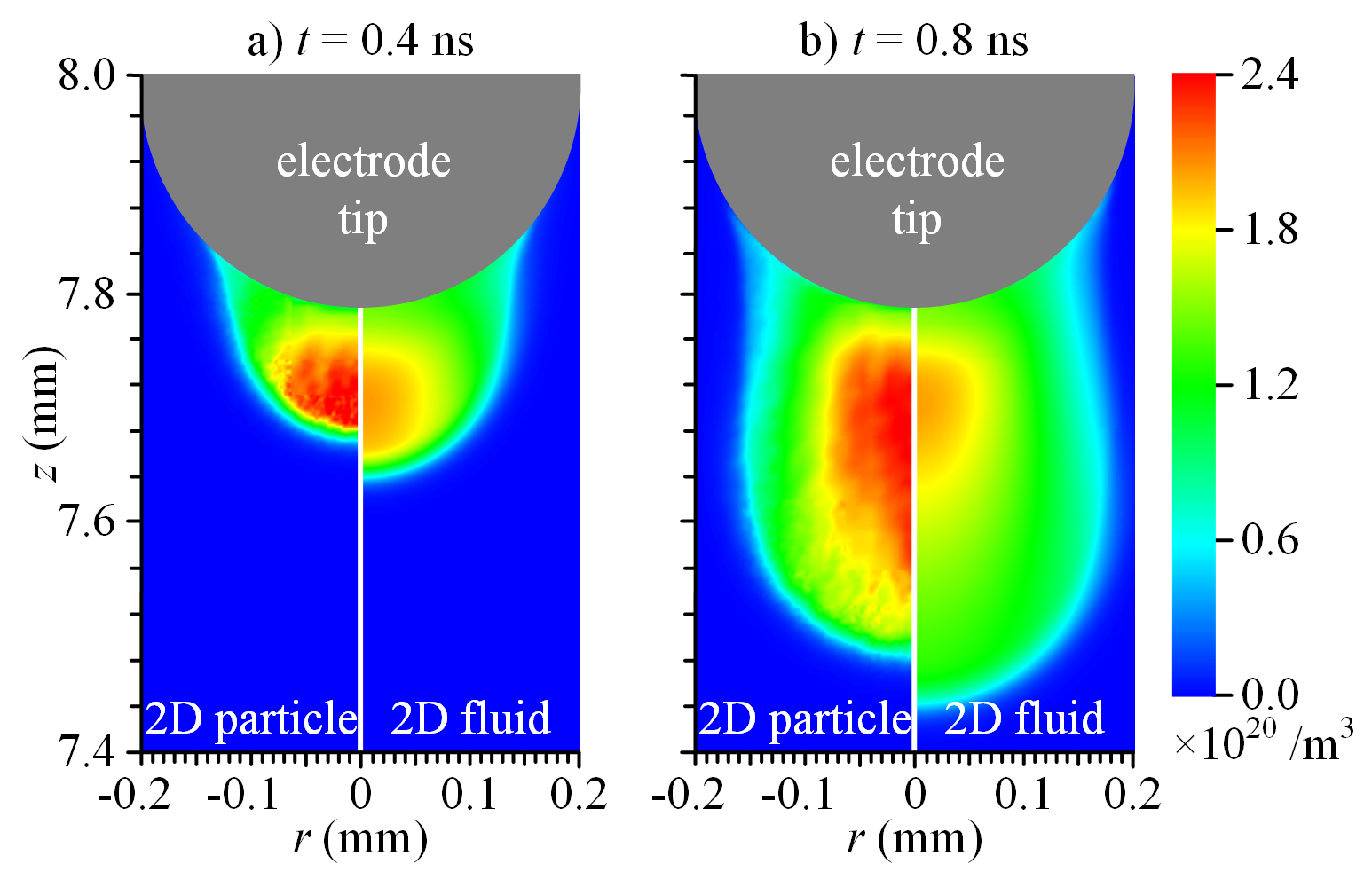}
    \caption{\revision{Electron density at 0.4 ns and 0.8 ns in the axisymmetric
        particle and fluid simulations, for an applied voltage of 11.70 kV.
        Results from one representative particle simulation are shown;
        stochastic fluctuations are initially small with the initial conditions
        used here.}}
    \label{fig:inception-comparison}
\end{figure}

For this study, we have designed the initial conditions such that inception
behavior would be similar in the particle and fluid simulations, by using a
sharp electrode and a compact initial seed with sufficiently many electrons. If
we define inception as the moment at which the streamer crosses the position
$z = 7.6 \, \textrm{mm}$, then inception is about 0.04 ns faster in the fluid
simulations. \revision{This is illustrated in figure
  \ref{fig:inception-comparison}, which shows the electron density at 0.4 ns and
  0.8 ns for the 2D fluid and particle simulations. The difference in streamer
  position at these times is primarily caused by faster inception in the fluid
  model. The difference increases somewhat in time, because a longer streamer
  propagates faster. Note that the electron density is higher in the particle
  model.}

% The main difference in streamer length during inception stage happens when electrons begin to multiply.

\revision{Faster inception in the fluid model} could be due to the local field approximation, with which
electrons are assumed to instantaneously relax to the background electric field.
Electron multiplication therefore happens more rapidly in the fluid simulations
at $t = 0$ ns, and similarly photoelectrons also instantaneously produce new
ionization. % On the
% other hand, energy relaxation is also quite fast in the particle model. In the
% absence of collisions an electron that is initially at rest gains an energy
% $\varepsilon$ in a time $\sqrt{2 m_e \varepsilon}/(e E)$. E.g., in a field of
% 100 kV/cm, it would only take about 0.3 ps to gain 1 eV.
We remark that when inception is highly stochastic (with different initial
conditions), another difference could be more relevant. With a fluid model low
densities always rapidly grow in a high field, even if they correspond to a
small probability of an electron being present, as was observed
in~\cite{Li_2021}. Such continuous growth of a low electron
density in high field regions can then lead to faster inception.
% We could maybe discuss the strong density gradients initially, but I'm not
% sure about this. Probably best to compare ionization rates once to find the
% answer to this question.

There is good agreement among the models for the streamer position versus
time, and thus also for the streamer velocities as a function of streamer
position. Velocity differences are generally less than 0.04 mm/ns among the
models. The mean relative deviation in velocity is below 2\%. We compute this quantity as
\begin{equation}
  \label{eq:velocity-diff}
  \left . \int |v_a(z) - v_b(z)| dz \middle/ \int v_a(z) dz\right . ,
\end{equation}
where $v_a$ and $v_b$ denote the velocities in the particle and fluid simulations, which are linearly interpolated between known positions.
After inception velocities increase approximately linearly with streamer
length. At the end of the gap they increase more rapidly due to boundary
effects.

\begin{figure}
    \centering
    \includegraphics[width=\linewidth]{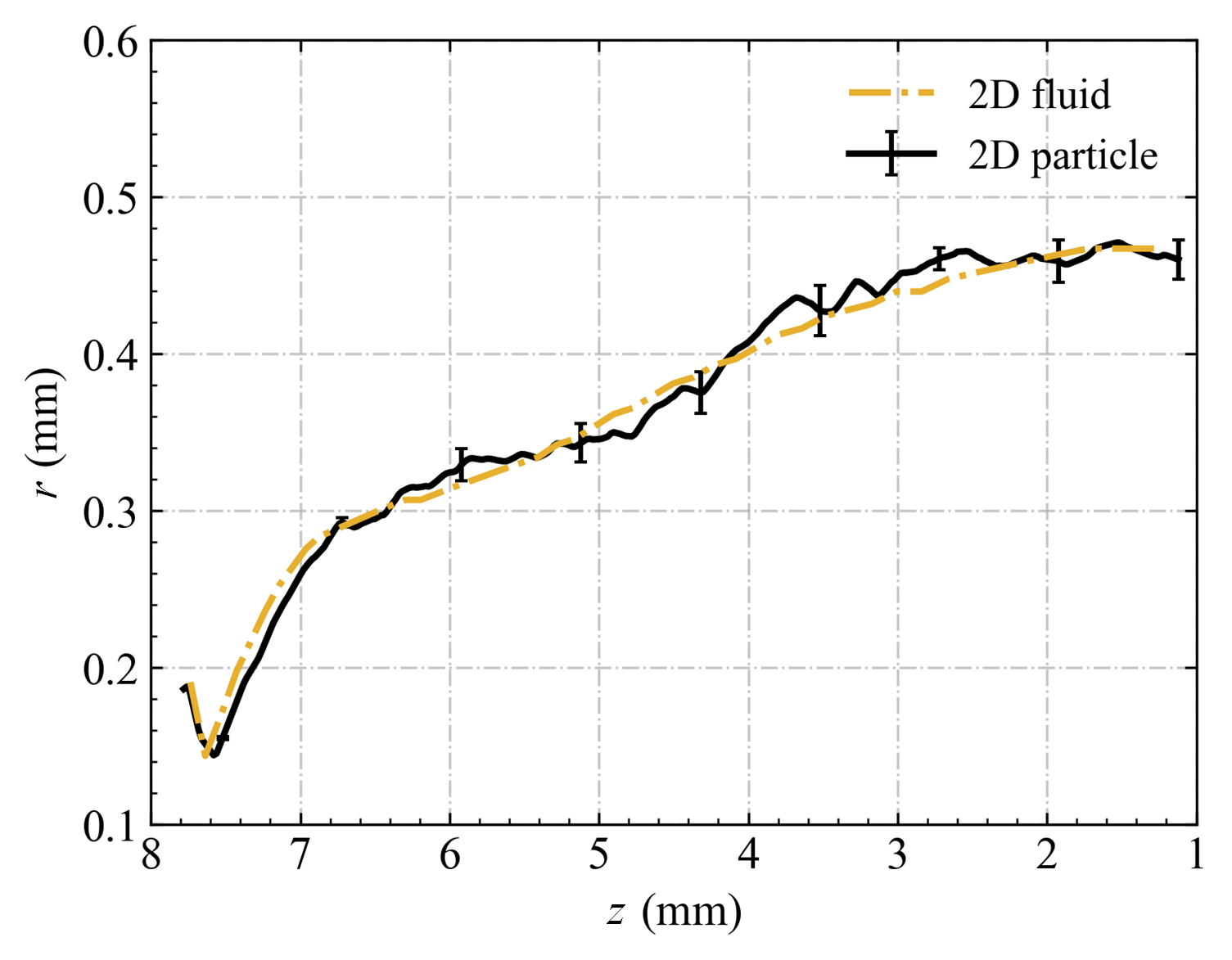}
    \caption{\revision{Streamer radius versus position for axisymmetric particle and fluid
      simulations at an applied voltage of 11.70 kV. The streamer radius was
      here defined as the radius at which $E_r$, the radial component of the
      electric field, is maximal. For the stochastic particle simulations the
      average of ten runs is shown, and the error bars indicate $\pm$ one
      standard deviation.}}
    \label{fig:radius-comparison}
\end{figure}

\revision{Figure \ref{fig:radius-comparison} shows the streamer radius versus
  position in axisymmetric particle and fluid simulations. Good agreement is
  found between the models, with the maximal difference in radius being below
  0.02 mm. Note that there are substantial fluctuations in the radius in the
  particle simulations as indicated by the error bars. For 3D simulations the
  radius is harder to compute, as it depends on the viewing angle. However, as
  can be seen from figure \ref{fig:15-results}, the radius appears to in good
  agreement between the 2D and 3D simulations.}

\subsection{Fluid model transport and reaction data}
\label{sec:TD-data}

\begin{figure*}
    \centering
    \includegraphics[width=\linewidth]{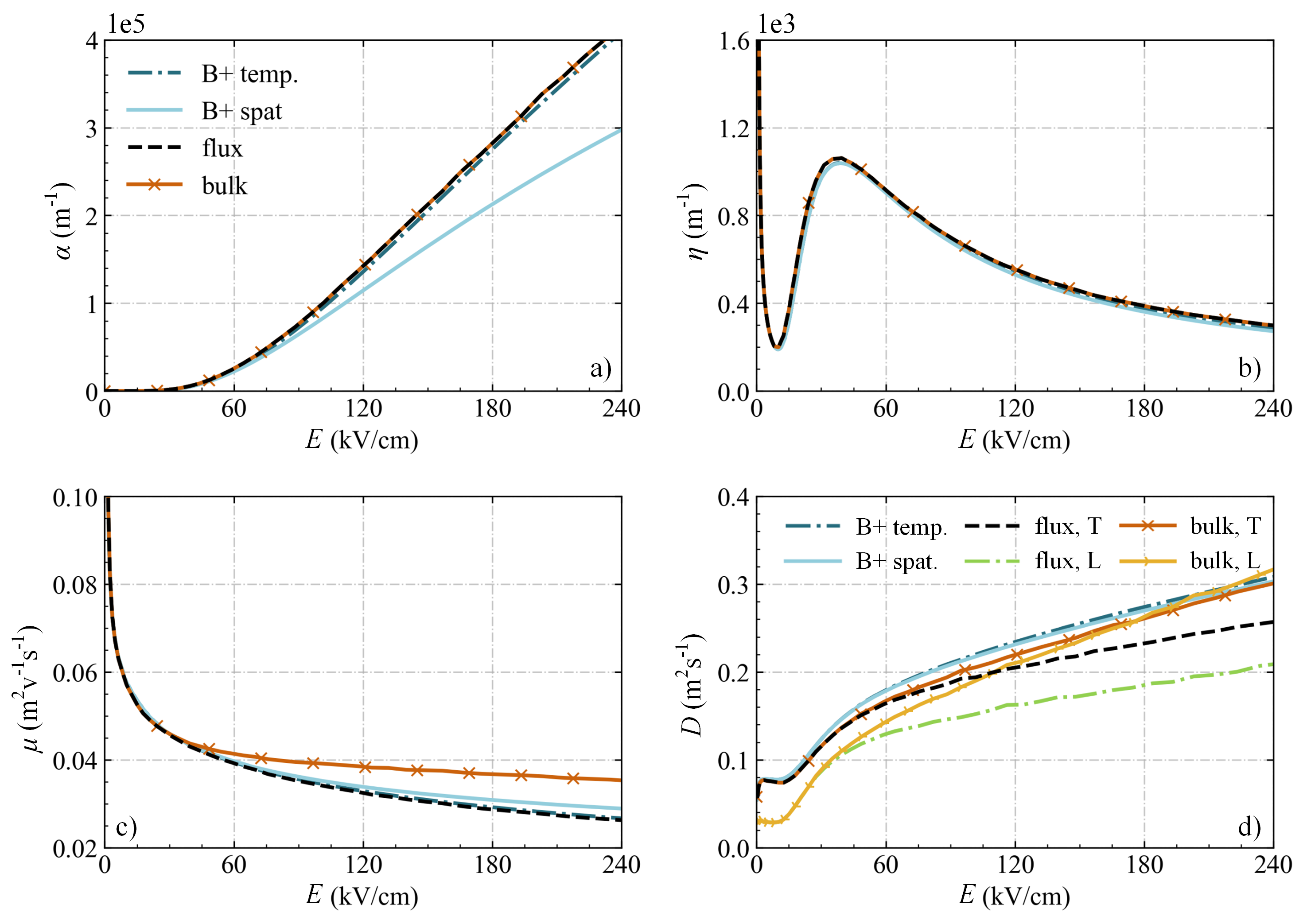}
    \caption{\revision{Electron transport data. a) Ionization
        coefficient, b) attachment coefficient, c) electron mobility and d)
        electron diffusion coefficient. The coefficients were computed for
        80$\%$ N$_2$ and 20$\%$ O$_2$ at 1 bar and 300 K, using Phelp's cross
        sections, see section \ref{sec:input-data}.} For BOLSIG+, data is shown
      using both a temporal and a spatial growth model. The data labeled
      ``bulk'' and ``flux'' was computed with a Monte Carlo swarm code. Both
      transverse and longitudinal diffusion coefficients were computed with this
      technique, but only the transverse coefficients are used in our fluid
      model.}
    \label{fig:transport_value_linear}
\end{figure*}

As mentioned in section \ref{subsec:fluid-input-data}, transport and reaction
data for a fluid model can be computed using different types of Boltzmann
solvers. Furthermore, both so-called \emph{flux} and \emph{bulk} data can be
computed. Flux data describes the behavior of individual electrons, whereas bulk
data describes the behavior of a group of electrons, taking ionization and
attachment into account. We here study how the choice of fluid model input data
affects the the consistency between particle and fluid simulations. The following types of input data are considered (with labels in bold):
\begin{itemize}
  \item (\textbf{B+ temp.}) Flux data computed with BOLSIG+ using its temporal growth model~\cite{hagelaar2005solving}. With this setting, the
  two-term approximation is solved by assuming that the electron density grows
  exponentially in time. This is the default growth model, but it is not clear
  whether it is the most suitable growth model for streamer
  simulations~\cite{hagelaar2005solving}.
  % (Here we chose non-Maxwellian EEDF and computed 400 quadratically-distributed data points of reduced electric field $E/N$, with a range of 1-1200 Td.
  \item (\textbf{B+ spat.}) Flux data computed with BOLSIG+ using its spatial growth
  model~\cite{hagelaar2005solving}, in which it is assumed that the electron
  density grows exponentially in space.
  \item (\textbf{flux}) Flux data computed with a Monte Carlo swarm
  method (available at ~\url{gitlab.com/MD-CWI-NL/particle_swarm}), which uses the same core routines for simulating electrons as our particle model~\cite{teunissen20163d}.
  \item (\textbf{bulk-a}) Bulk data computed with the same Monte Carlo swarm
  method. In this variant, only the transport terms in equation
  (\ref{eq:fluid-model}) are modified, by computing the electron flux as
  $-n_e\mu_{e}^B \mathbf{E} - D_{e}^B \nabla n_e$, where $\mu_{e}^B$ and
  $D_{e}^B$ denote bulk coefficients.
  \item (\textbf{bulk-b}) The same bulk data as above, but in this variant the
  reaction terms in equation (\ref{eq:fluid-model}) are also modified by
  multiplying them with $\mu_{e}^B / \mu_{e}$, where $\mu_{e}$ denotes the
  standard flux mobility.
\end{itemize}
With the \textbf{bulk-a} approach reaction rates are the same as with flux data.
However, the number of reactions taking place per unit length (traveled by
electrons) is changed, i.e., the so-called Townsend coefficients are different.
With the \textbf{bulk-b} approach it is the other way around.

Different types of transport data are shown in figure
\ref{fig:transport_value_linear}. Above about 180 Td ionization becomes
important and bulk mobilities are larger than flux mobilities. The spatial
growth model of BOLSIG+ leads to a significantly smaller ionization coefficient.
In high electric fields, its value is about 25-30\% less than that of the other
approaches. With the Monte Carlo approach both transverse and longitudinal diffusion
coefficients are computed, but in our fluid simulations we for simplicity only
use the transverse ones. The BOLSIG+ flux diffusion coefficient also corresponds
to the transverse direction~\cite{Pitchford_1982,hagelaar2005solving}, but it is larger than the Monte Carlo flux coefficient. Such differences between diffusion coefficients computed with a two-term approach and higher-order methods have been observed before, see e.g.~\cite{Petrovic_2009a}. However, the different diffusion coefficients only have a minor impact on our simulations, as shown below.

\begin{figure}
    \centering
    \includegraphics[width=\linewidth]{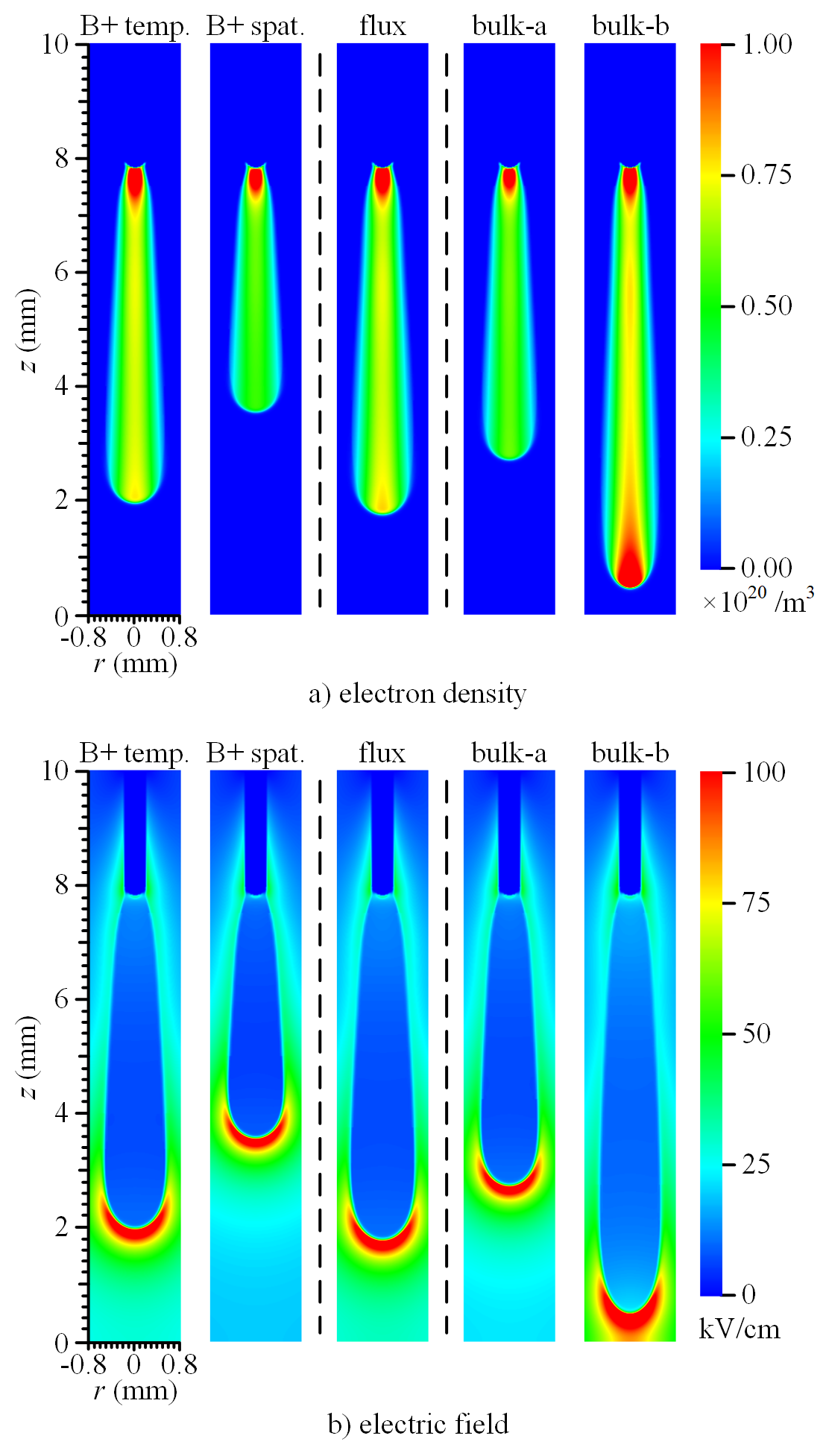}
    \caption{Electron densities and electric fields at $t$ = 9.6 ns for axisymmetric fluid simulations with an applied voltage of 11.7 kV. Different types of transport data are used, from left to right: BOLSIG+ with temporal growth, BOLSIG+ with spatial growth, Monte Carlo flux data, and two types of Monte Carlo bulk data. With bulk-a only transport terms are modified, and with bulk-b reaction terms are also scaled with the bulk mobility.}
    \label{fig:fluid-result}
\end{figure}

\begin{figure}
  \centering
  \includegraphics[width=\linewidth]{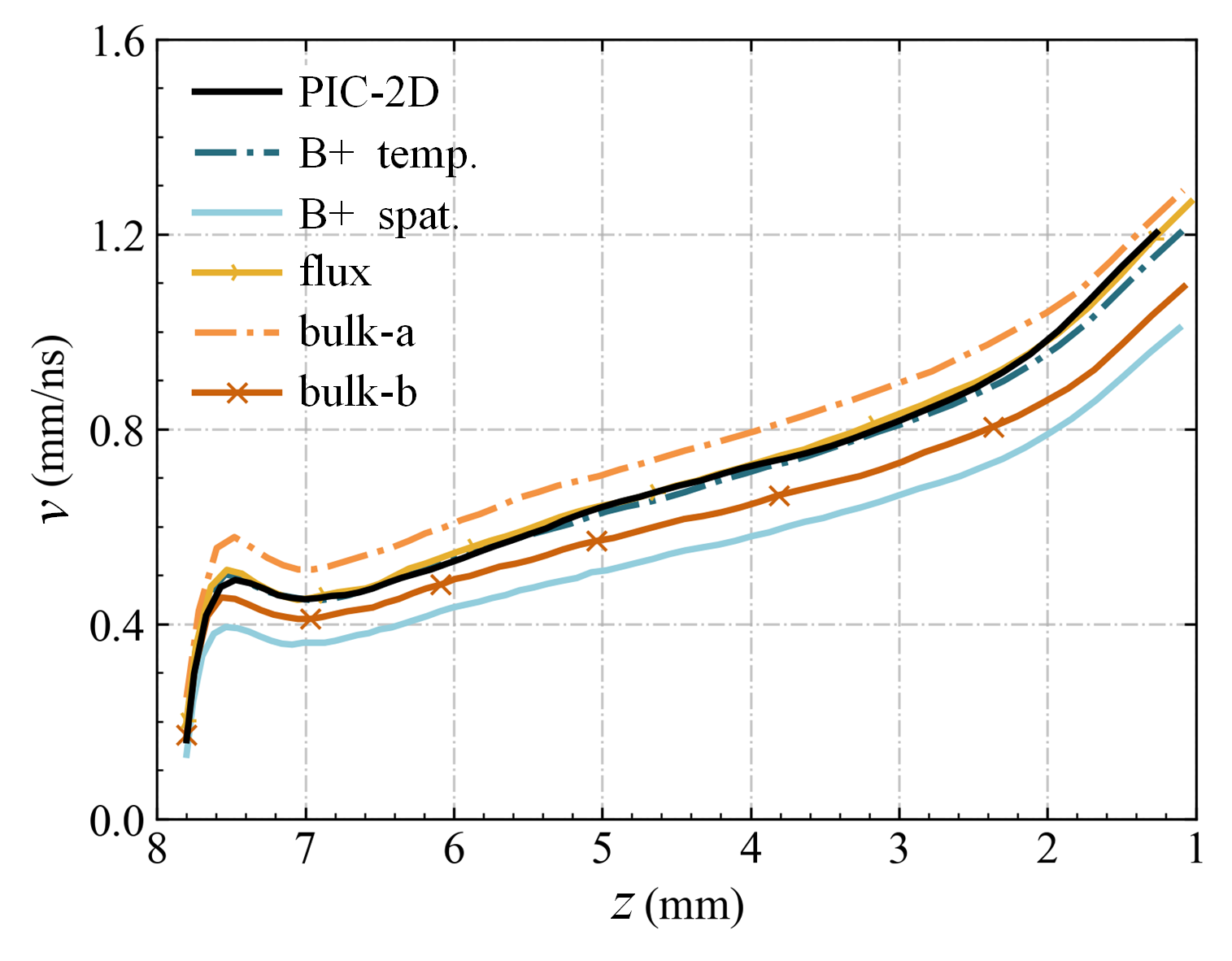}
  \caption{Streamer velocity versus streamer head position for different types
    of transport data. The labels are explained in figure \ref{fig:fluid-result}
    and in section~\ref{sec:TD-data}.}
  \label{fig:fluid-comparison}
\end{figure}

Figure \ref{fig:fluid-result} shows axisymmetric fluid simulations with the
input data listed above; streamer positions over time are given in
table~\ref{tab:head-position} and streamer velocities in figure
\ref{fig:fluid-comparison}.
There are minor differences in streamer velocity when comparing the BOLSIG+ flux data with temporal growth and the Monte Carlo flux data. When comparing streamer velocities at the same length, relative differences are below 3\%. % \jt{3.3~\%}.
With both types of data good agreement is obtained with the axisymmetric particle simulations.
% The difference between two cases with different flux data is within .
In contrast, the BOLSIG+ data with the spatial growth model leads to a streamer velocity that is much too low, due to the lower ionization coefficient.

Both types of bulk transport data lead to significant deviations compared to the particle model. When only the transport coefficients are changed (bulk-a), the streamer is significantly slower and it has a lower degree of ionization. With this data electrons drift faster, but the degree of ionization produced in the streamer channel is lower, leading to a slower discharge.
However, when the reaction terms are also changed (bulk-b), the streamer propagates too fast. The higher streamer velocity is to be expected, since most terms on the right-hand side of equation~(\ref{eq:fluid-model}) are now scaled with the bulk mobility.

In conclusion, bulk data and data computed with a spatial growth model are not
recommended for the simulation of positive streamers. With flux transport data
there are minor differences between BOLSIG+ data computed with a temporal growth
model and Monte Carlo data, but both lead to good
agreement with the particle simulations.
% For both bulk transport data cases, the maximal value of velocity difference compared to flux transport data from swarm experiment case can reach 0.2 mm/ns.

\subsection{Mesh refinement and numerical convergence}
\label{sec:mesh-refin-numer}

\begin{figure}[h]
  \centering
  \includegraphics[width=\linewidth]{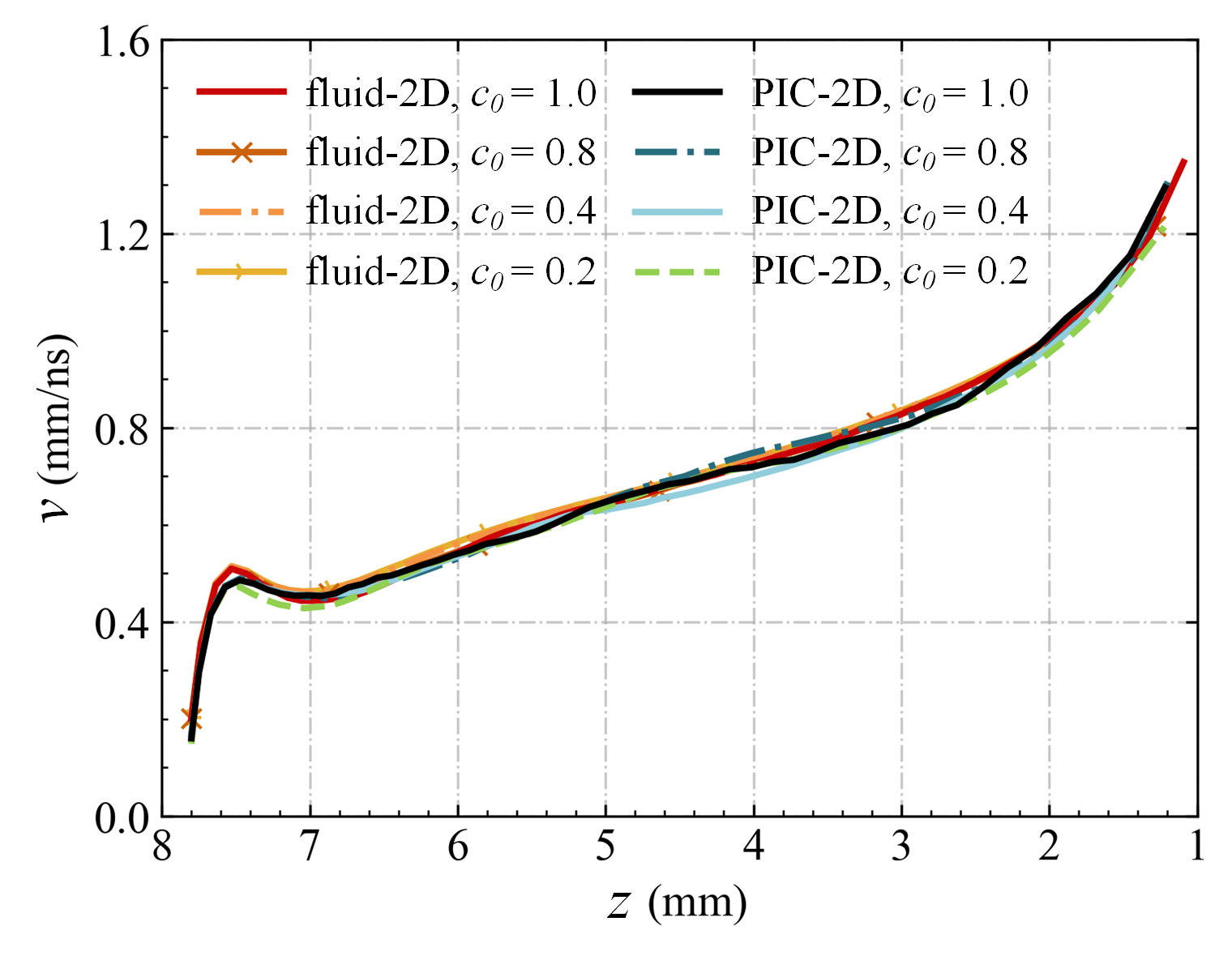}
  \caption{Streamer front velocities versus streamer head position for different
    refinement criteria. Other conditions are the same as in section
    \ref{subsec:15-comparison}. For the particle simulations the average of ten
    runs is shown to reduce stochastic fluctuations.}
  \label{fig:adx}
\end{figure}

\begin{table}
  \centering
  \begin{tabular}{lllll}
    Model      & $c_0$     & $z$ (3 ns) & $z$ (6 ns) & $z$ (9 ns)   \\
    \hline
    PIC-2D     & 0.2       & 6.51    & 4.81    & 2.59    \\
               &           & $\pm 0.02$    & $\pm 0.09$    & $\pm 0.11$    \\
    PIC-2D     & 0.4       & 6.45    & 4.74    & 2.54    \\
               &           & $\pm 0.03$    & $\pm 0.04$    & $\pm 0.06$    \\
    PIC-2D     & 0.8       & 6.45    & 4.74   & 2.41    \\
               &           & $\pm 0.03$    & $\pm 0.11$    & $\pm 0.21$    \\
    PIC-2D     & 1.0       & 6.45    & 4.71    & 2.45    \\
               &           & $\pm 0.04$    & $\pm 0.13$    & $\pm 0.13$    \\
    \hline
    fluid-2D   & 0.2       & 6.37    & 4.54    & 2.14    \\
    fluid-2D   & 0.4       & 6.37    & 4.55    & 2.17    \\
    fluid-2D   & 0.8       & 6.40    & 4.64    & 2.31    \\
    fluid-2D   & 1.0       & 6.42    & 4.67    & 2.35
    % fluid-3D & BOLSIG+ ( & \jt{TODO?} & \jt{TODO?} & \jt{TODO?} \\
    % fluid-3D & BOLSIG+   & \jt{TODO?} & \jt{TODO?} & \jt{TODO?} \\
  \end{tabular}
  \caption{Streamer head position ($z$, in mm) at 3, 6 and 9 ns, using an applied voltage of 11.7 kV. Different values of the refinement parameter $c_0$ are used, see section \ref{sec:afivo-amr-framework}. For the particle simulations averages over ten runs are shown, together with the standard deviation of the sample. Streamer lengths are given by $7.8 \, \textrm{mm} - z$.}
  \label{tab:streamer-pos}
\end{table}

We here study the sensitivity of the particle and fluid simulations to the grid spacing, to test whether our simulations are close to numerical convergence. To control the grid spacing, the refinement parameter $c_0$ is varied, see section \ref{sec:afivo-amr-framework}.
Note that the time step in both models will also be affected by the grid spacing, as explained in section \ref{sec:time-step-pic} and \ref{sec:time-step-fluid}.

Figure \ref{fig:adx} shows streamer velocities versus streamer position for
$c_0$ values of 1.0, 0.8, 0.4 and 0.2, for which the minimal grid spacing is
3.9, 3.9, 1.9, and 0.9 $\mu$m, respectively. Streamer positions at $t$ = 3, 6
and 9 ns are given in table \ref{tab:streamer-pos}. With the fluid model,
deviations in length at $t = 9$ ns are about 3\% with $c_0 = 0.8$ compared to
the finest-grid case. With the particle model, there are statistical
fluctuations that make it harder to establish numerical convergence, but at
$t = 9$ ns streamer lengths are also within 3\% for all tested cases. When
comparing the streamer velocity versus position for $c_0 = 0.8$ and $c_0 = 0.2$,
convergence errors are about 1\% for the fluid model and about 2\% for
the particle model, using equation (\ref{eq:velocity-diff}).

Table \ref{tab:streamer-pos} allows to compare differences in streamer length
between particle and fluid simulations using the same refinement. Interestingly, these
differences are larger on finer grids: with $c_0 = 0.2$, the relative differences in streamer length are about 8--10\% at 3, 6 and 9 ns, whereas for $c_0 = 0.8$ they are about 2--4\%. For streamer velocities
(compared at the same streamer length) the mean deviations are about
4\% and 2\% for these two cases, using equation (\ref{eq:velocity-diff}).

Based on the above, we conclude that numerical convergence errors are relatively
small for our default refinement parameter $c_0 = 0.8$ -- they do at least not
exceed the intrinsic differences between the models, which are already quite
small. For the test case considered here, with an applied voltage of 11.7 kV,
the difference in streamer velocity is about twice as large (4\% instead of 2\%)
on the finest grid. The main reason for this is that in the fluid simulations,
which are more sensitive to the grid refinement, the streamer velocity is
somewhat higher on finer grids. In section \ref{sec:results-at-different} we
show that for higher applied voltages, the velocity is actually higher in the
particle simulations. We therefore expect that using a finer grid somewhat increases model discrepancies for lower applied voltages, and that is somewhat reduces model discrepancies for higher applied voltages.

% Velocity differences: less than 8.1~\%, 5.1~\%,
% 3.6~\% and 4.7~\% for $c_0 = 0.2, 0.4, 0.8$ and $1.0$.

% Differences in streamer velocity compared at the same position were less than
% 6.45~\% for the particle simulations and 6.48~\% for the fluid simulations.

% less than 0.18 mm/ns and 0.22 mm/ns for axisymmetric particle and fluid models, respectively. When comparing these velocities between the particle and fluid models, the discrepancies of streamer velocity for $c_0$ = 0.2,0.4 and 0.8 cases are less than 0.34, 0.26 and 0.11 mm/ns, which means the difference between models slightly increase under finer grids, but among all cases two axisymmetric models show acceptable numerical discrepancies.

\subsection{Stochastic fluctuations}
\label{sec:role-stoch-effects}

\begin{figure*}
    \centering
    \includegraphics[width=\linewidth]{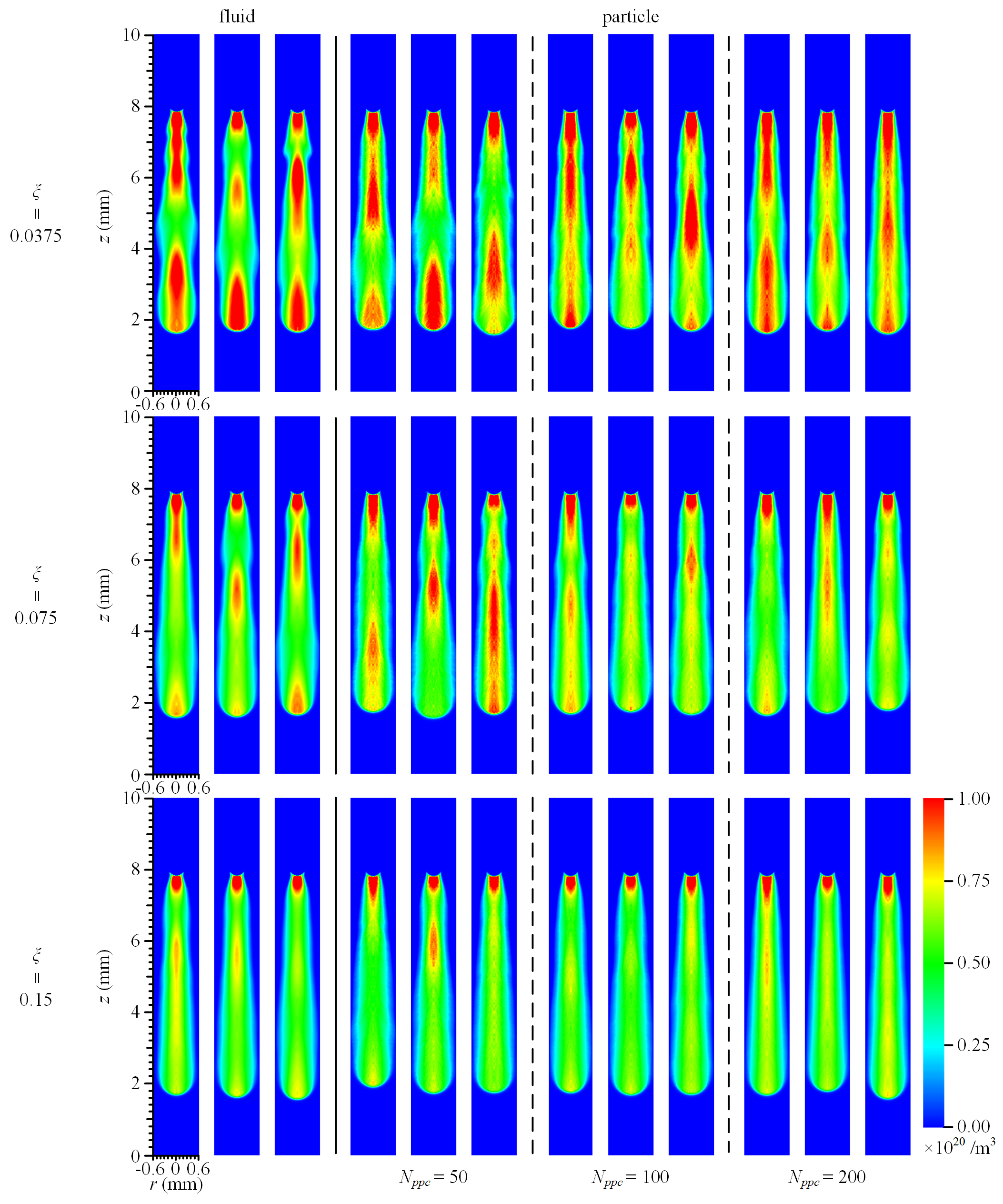}
    \caption{Electron densities at $t$ = 10 ns in axisymmetric fluid and particle simulations. The photoionization coefficient $\xi$ and the desired number of particles per cell $N_{ppc}$ are varied. For each combination of two parameters, three runs are shown. Stochastic photoionization is now also used in the fluid model. The condition are otherwise the same as in section 3.1.}
    \label{fig:fluctuation}
\end{figure*}

To investigate the source of stochastic fluctuations in the axisymmetric simulations we vary
two parameters. The first is the photoionization factor $\xi$, see equation
\ref{eq:UV-source-term}, which is a proportionality factor that relates the
number of UV photons produced to the electron impact ionization source term. It
therefore directly controls the amount of photoionization. To study how $\xi$
affects stochastic behavior, we use the discrete photoionization model in
both the particle and the fluid simulations presented here. The second parameter we vary is $N_{ppc}$, which controls the `desired' number of particles per cell in particle simulations, see equation \ref{eq:weight-calculation-superparticle}.

Figure \ref{fig:fluctuation} shows results of axisymmetric particle and fluid models for $\xi$ = 0.0375, 0.075, 0.15 and $N_{ppc}$ = 50, 100, 200.
In both models the streamer length is not sensitive to the amount of photoionization, as was also observed in e.g.~\cite{wormeester2010probing}. However, fluctuations in the electron density are significantly larger for the $\xi$ = 0.0375 case, whereas these fluctuations are reduced for the $\xi$ = 0.15 case, as was also observed in \cite{bagheri2019effect}. With $\xi$ = 0.0375 we even observed branching in a few of the simulation runs, which is probably due to increased density fluctuations near the $z$-axis when the amount of photoionization is decreased. Fluctuations in the streamer radius are also larger for a lower value of $\xi$.
When $N_{ppc}$ is increased, fluctuations in electron densities and streamer radius are slightly reduced, but the effect is weaker than that of the $\xi$ parameter.
We therefore conclude that the discrete photoionization model is responsible for most of the stochastic fluctuations in our results. This confirms the assumptions made in recent work~\cite{bagheri2019effect,Marskar_2020}, in which fluid models were used to demonstrate the importance of stochastic photoionization on streamer branching.

Finally, we remark that in figure \ref{fig:fluctuation} the stochastic
fluctuations are demonstrated with axisymmetric models, in which these
fluctuations are not completely physical. We have also performed 3D fluid
simulations with stochastic photoionization, in which these fluctuations looked
qualitatively similar to those shown in figure \ref{fig:15-results} for the 3D
particle model.
% We could show this with a figure for the revision
However, a statistical comparison of these 3D models for the
parameter range shown in figure \ref{fig:fluctuation} could not be performed due
to the high computational costs of the 3D particle simulations.

\subsection{Results at different voltages}
\label{sec:results-at-different}

\begin{figure}
    \centering
    \includegraphics[width=\linewidth]{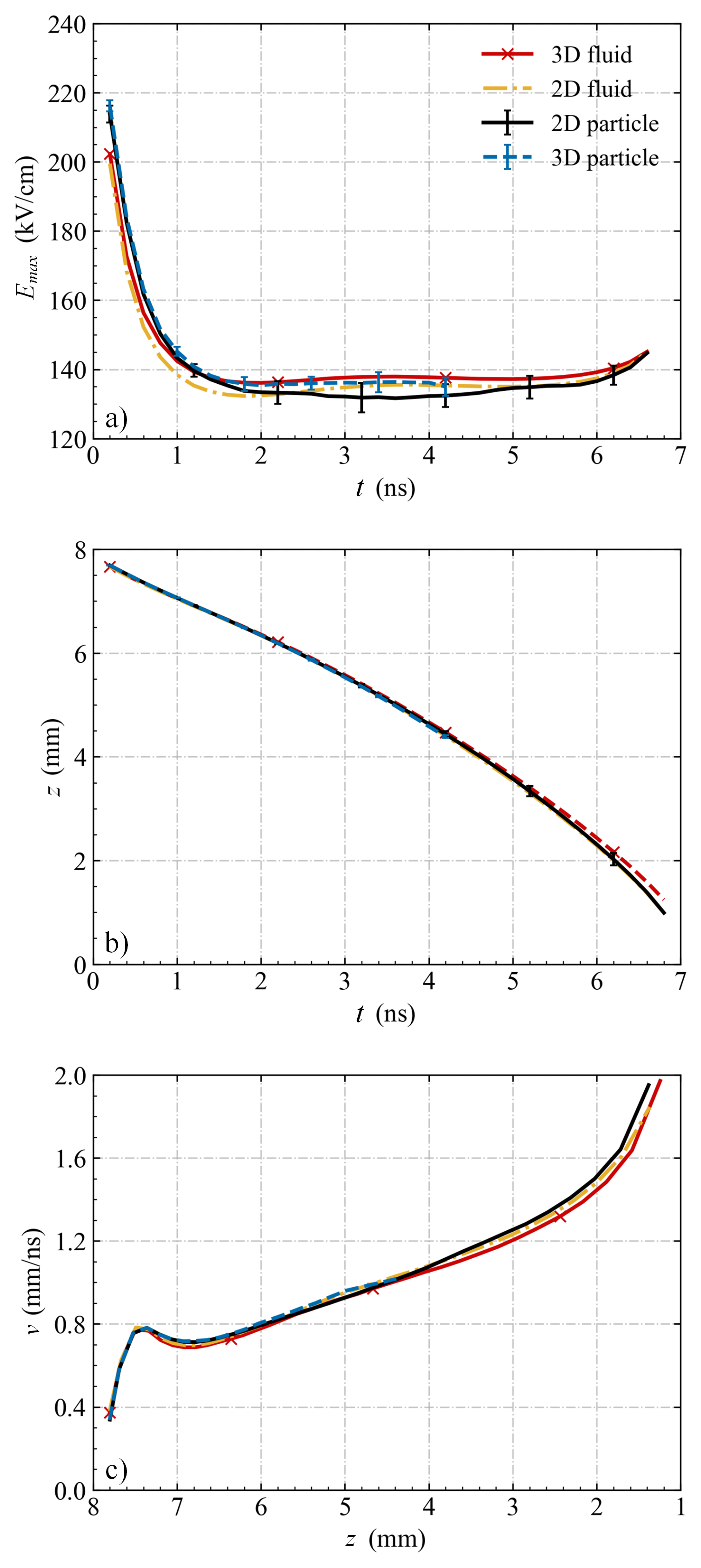}
    \caption{Comparison between axisymmetric and 3D particle and fluid models at
      an applied voltage of 14.04 kV, similar to figure~\ref{fig:15-comparison}.
      From top to bottom: the maximal electric field and streamer position
      versus time, and streamer velocity versus streamer position. For the
      particle model the average of ten runs is shown with error bars indicating
      $\pm$ one standard derivation.}
    \label{fig:18-comparison}
\end{figure}

Figure \ref{fig:18-comparison} shows results for particle and fluid simulations
at a higher applied voltage of 14.04 kV, which results in a background electric
field of about 18 kV/cm. All the other parameters are the same as in section
\ref{subsec:15-comparison}. For the 3D particle model results at later times are
missing, because these simulations exceeded the memory and time constraints of
our computational hardware, see \ref{sec:computational-costs}.

At this higher voltage, the agreement between the models is of similar quality
as in figure \ref{fig:15-comparison}, but there are a few differences. Figure
\ref{fig:18-comparison}(a) shows that inception is significantly faster. The
relaxation of the initial high field takes place in about 1 ns, so roughly twice
as fast, and the curves for the maximal electric field are now in better
agreement. With a higher applied voltage the streamer velocity is higher, but
the propagation is otherwise similar to that in figure~\ref{fig:15-comparison}.
The agreement between the models is still good: between the 2D particle and 2D
fluid simulations, the mean deviation in velocity (compared at the same streamer
length) is about 1\%. However, the discrepancy between the 2D and 3D fluid
simulations is now somewhat larger. This is probably due to the difference in
computational domains and electrostatic boundary conditions in 2D and 3D, which
could play a stronger role for a more conducting streamer channel at a higher
voltage. The sensitivity of discharge simulations to these boundary conditions
was recently observed in \cite{Li_2021}.

\begin{figure}
    \centering
    \includegraphics[width=\linewidth]{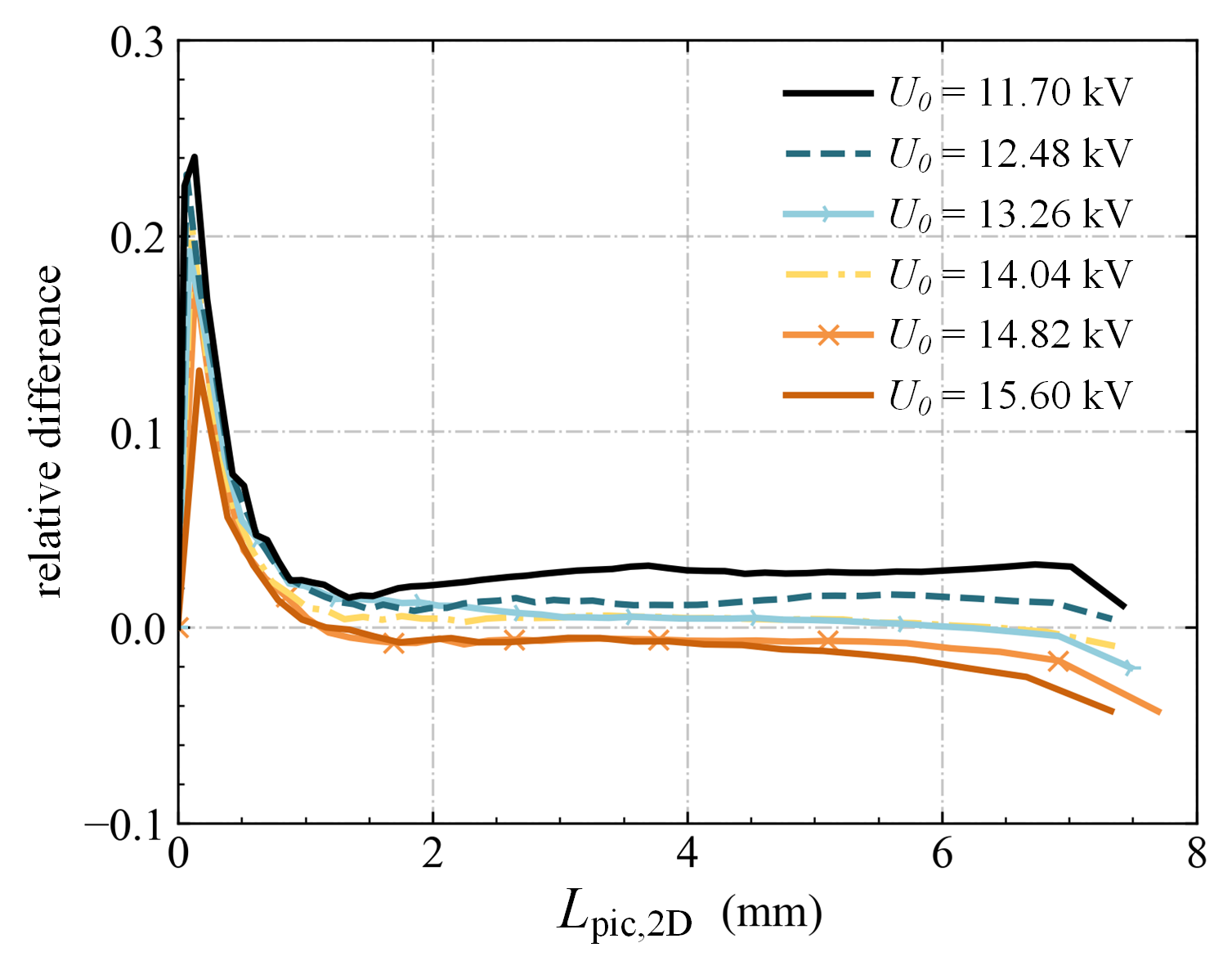}
    \caption{The relative difference $\Delta_L$ in streamer length versus time,
      shown for axisymmetric particle and fluid simulations at several voltages.
      The difference is computed as
      $\Delta_L = (L_\mathrm{fluid,2D} - L_\mathrm{pic,2D}) /
      L_\mathrm{pic,2D}$. Besides the applied voltage, simulations conditions
      are the same as in section \ref{subsec:15-comparison}.}
    \label{fig:multi-voltage}
\end{figure}

Figure \ref{fig:multi-voltage} shows the relative difference $\Delta_L$ in
streamer length between axisymmetric particle and fluid models for applied
voltages from 11.70~kV to 15.60~kV. These voltages correspond to background
electric fields of about 15~kV/cm to 20~kV/cm. The difference is computed as
\begin{equation*}
  \Delta_L = (L_\mathrm{fluid,2D} - L_\mathrm{pic,2D}) / L_\mathrm{pic,2D},
\end{equation*}
where $L_\mathrm{fluid,2D}$ and $L_\mathrm{pic,2D}$ are the streamer lengths in the fluid and particle model at a particular time.

In all cases, $\Delta_L$ peaks during
streamer inception. This happens because inception occurs faster in the fluid
model, as explained in section~\ref{subsec:15-comparison}, and because the
denominator is initially small. For higher applied voltages, the initial peak in $\Delta_L$ becomes smaller.

In which model the streamer has advanced the furthest at a particular time
depends on the applied voltage. When $U_0$ is lower than 14.04 kV, $\Delta_L$ is
generally positive, whereas for higher voltages it becomes negative. This
indicates that relative to the fluid simulations, the velocity in the particle
simulations is higher at higher voltages. The mean deviations between
velocities in the particle and fluid simulations are 1.8\% (11.7 kV case), 1.9\%
(12.48 kV), 1.3\% (13.26 kV), 1.3\% (14.04 kV), 1.4\% (14.82 kV) and 2.4\% (15.6
kV).

\section{Conclusions}
\label{sec:conclusions}

We have quantitatively compared a PIC-MCC (particle-in-cell, Monte Carlo
collision) model and a drift-diffusion-reaction fluid model with the local field
approximation for simulating positive streamer discharges. The simulations were
performed in air at 1 bar and 300 K, in background fields below breakdown ranging from 15 kV/cm to 20
kV/cm, using both axisymmetric and fully three-dimensional geometries.

We have found surprisingly good agreement between the particle and fluid
simulations. Streamer properties such as maximal field, radius, and velocity
were all very similar. When compared at the same streamer length, the mean
difference in streamer velocity was generally below 4\%. One source of
differences was the photoionization model, for which we used a stochastic
approach in the particle simulations and a continuum approach in most of the
fluid simulations.

% compared streamer velocities at the same streamer length, and found differences
% below 8\%.

We have investigated the effect of different types of transport data in fluid
models, how well the models are numerically converged, what the main source of
stochastic fluctuations is, and how the agreement between the models is affected
by the applied voltage. Our main conclusions on these topics are:
\begin{itemize}
  \item The type of transport data used in a fluid model is important. By using
  flux transport coefficients computed with a Monte Carlo approach or BOLSIG+
  (using its temporal growth model), good agreement is obtained between the
  fluid and particle simulations. The use of bulk coefficients leads to either
  faster or slower streamer propagation, depending on how the coefficients are
  used. Data computed with the spatial growth model of BOLSIG+ leads to a significantly slower streamer discharge.
  \item Numerical convergence errors are small in the particle and fluid
  simulations presented here. We have compared axisymmetric particle and fluid
  simulations with grid refinement satisfying $\alpha(E) \Delta x < c_0$ for
  $c_0 = 0.2, 0.4, 0.8$ and $1.0$, where $\alpha(E)$ is the ionization
  coefficient. For an applied voltage of 11.7 kV, convergence errors in streamer
  velocity (compared at the same position) were about 1\% for the fluid
  simulations and about 2\% for the particle simulations. On the finest grids,
  streamer velocities increased slightly in the fluid simulations.
  \item Stochastic fluctuations are visible in axisymmetric and 3D particle
  simulations, for example in the streamer's degree of ionization, maximal
  electric field and radius. In our simulations, the dominant source of these
  stochastic fluctuations is discrete photoionization. Fluid simulations with
  the same discrete photoionization model exhibit similar fluctuations as
  particle simulations. Due to these fluctuations, streamers in 3D simulations
  propagate slightly off-axis. In the particle simulations, the number of
  particles per cell did not significantly affect these fluctuations.
  \item Axisymmetric simulations were performed for applied voltages between
  11.7 kV to 15.6 kV, corresponding to background fields of about 15 kV/cm to 20
  kV/cm. Discrepancies in streamer length (versus time) between particle and
  fluid simulations were generally below 3\%. The mean deviations in streamer
  velocity (versus length) were about 2\% for all applied voltages. Other
  streamer properties, such as the maximal electric field, were also in good
  agreement.
\end{itemize}

\revision{Finally, we expect differences between particle and fluid models to
  increase at lower applied electric fields. Inception will be more stochastic,
  and a smaller streamer radius in lower fields will lead to steeper gradients
  in the electron density and electric field, which could increase errors due to
  the local field approximation. However, a comparison in lower fields, in which
  streamer branching would probably have to be taken into account, is left for
  future work.}
\section*{Acknowledgments}

Part of this work was carried out on the Dutch national e-infrastructure with the support of SURF Cooperative. This work was partly funded by the China Scholarship Council (CSC) (Grant No. 202006280465). This work was supported by the National Natural Science Foundation of China (Grant No. 51777164).

\section*{Data availability statement}

The data that support the findings of this study are openly available at the following URL/DOI: \url{https://doi.org/10.5281/zenodo.5509678}.

\appendix

\section{Computational cost}
\label{sec:computational-costs}

Typical computing times for the results in section \ref{subsec:15-comparison} were a few minutes (2D fluid model), 8h (3D fluid model), and 2-3h (2D particle model). These computations ran on an Intel(R) Core(TM) i9-9900K eight-core processor, using OpenMP parallelization. The 3D particle simulations were performed on Cartesius, the Dutch national supercomputer. A single thin node with a Intel Xeon E5-2690 v3 (Haswell) 24-core processor and 64 GB of RAM was used. Computations ran for up to five days, with up to 600 million particles.

\revision{The maximum number of grid cells used for the simulations presented in section \ref{subsec:15-comparison} were: $ 4.2\times10^4$ (2D fluid), $1.4\times10^8$ (3D fluid), $ 5.1\times10^4$ (2D particle), $ 8.3\times10^7$ (3D particle). For the numerical convergence tests presented in section \ref{sec:mesh-refin-numer}, the maximum number of grid cells at $t = 9$ ns were $ 2.4\times10^5$, $ 8.5\times10^4$, $ 3.8\times10^4$, $ 3.3\times10^4$ (2D fluid, $c_0 = 0.2, 0.4, 0.8, 1.0$ respectively) and $ 2.0\times10^5$, $ 8.2\times10^4$, $ 4.2\times10^4$, $ 3.5\times10^4$ (2D particle, $c_0 = 0.2, 0.4, 0.8, 1.0$ respectively).}

\section*{References}.
\bibliographystyle{unsrt}
\bibliography{comparison-ref}

\end{document}